%% file: paper.tex
\documentclass{article}

\usepackage{booktabs}       
\usepackage{arxiv}
\usepackage{float}
\usepackage[usenames,dvipsnames]{color}
\usepackage{placeins}
\usepackage{multirow}
\usepackage{orcidlink}
\usepackage{xspace}

\RequirePackage{uclthesis}

\usepackage[sorting=none,style=numeric-comp]{biblatex}
\renewbibmacro*{in:}{%
  \iffieldundef{journaltitle}
    {}
    {\printtext{\bibstring{in}\intitlepunct}}}

\newcommand{\printtitle}{Better Queries, Cheaper Attention: Adapting Transformers for Efficient Sparse Reconstruction}

\title{\printtitle}

\author{
Philippa Duckett\,\orcidlink{0000-0002-8893-3512}\textsuperscript{1},
Samuel Van Stroud\,\orcidlink{0000-0002-7969-0301}\textsuperscript{1},
Max Hart\,\orcidlink{0009-0004-5309-911X}\textsuperscript{1},
Gabriel Facini\,\orcidlink{0000-0002-4056-4578}\textsuperscript{1},
and Tim Scanlon\,\orcidlink{0000-0002-2746-525X}\textsuperscript{1}
\\
\textsuperscript{1}\textit{Centre for Data Intensive Science and Industry, Department of Physics and Astronomy, University College London},
}
\renewcommand{\shorttitle}{Adapting Transformers for Efficient Sparse Reconstruction}

\hypersetup{
    colorlinks,
    pdftitle={Adapting Transformers for Efficient Sparse Reconstruction},
    pdfsubject={},
    pdfauthor={Philippa Duckett et al},
    pdfkeywords={Particle Physics, ML Reconstruction, Object Detection},
    citecolor=blue,
    urlcolor=blue,
    linkcolor=blue,
}

\begin{document}

\newcommand{\ours}{\textsc{Name}\xspace}

\maketitle

\vspace{8em}
\begin{abstract}

Query-based transformer decoders are effective for object reconstruction from sparse scientific sensor measurements, but their scalability to high-multiplicity data is limited by fixed, input-independent query sets and costly decoder cross-attention. We introduce a geometry-aware dynamic-query decoder that couples input-conditioned query construction with structured sparse cross-attention. Decoder queries are initialised from selected encoder-level measurement representations that serve as candidate trajectory seeds, making both query content and query multiplicity input-dependent. Local Strided Cross-Attention (LSCA) exploits the induced geometric ordering by replacing learned mask-gated cross-attention with a geometry-defined local support that restricts attention to physically plausible query--hit interactions and exposes sparsity for block-sparse execution. We study this architecture for charged-particle trajectory reconstruction in a simplified High-Luminosity Large Hadron Collider detector, where thousands of trajectories must be reconstructed from tens of thousands of sparse measurements. In the nominal configuration, the dynamic-query (DQ) architecture increases trajectory reconstruction efficiency from $94.1\%$ to $98.1\%$ and reduces the fake rate by more than a factor of two relative to the fixed-query baseline. The DQ+LSCA model reduces end-to-end inference latency by nearly $50\%$ and peak allocated inference memory by more than a factor of $10$ relative to the fixed-query baseline.

\end{abstract}
\vfill


\newpage
\input{sections/1.introduction}
\input{sections/2.related}
\input{sections/3.HEP}

\input{sections/4.model}
\input{sections/5.experiments}
\input{sections/6.results}
\input{sections/7.conclusion}

\section*{Acknowledgments}
We thank Nikita Pond, Sunny Wong, Wei Sheng Lai, Edoardo Critelli, Jackson Barr, and S\'ebastien Rettie for fruitful discussions. We also extend our thanks to UCL for the use of their high-performance computing facilities, with special thanks to Edward Edmondson for his expert management and technical support.

We gratefully acknowledge the support of the UK's Science and Technology Facilities Council (STFC), UCL Centre for Doctoral Training in Data Intensive Science, and UK Research and Innovation (UKRI) via ST/X005992/1, ST/W00674X/1, ST/W00058X/1, and UKRI3900. P.D. and M.H. are also supported by departmental and industry contributions and T.S. is supported by the Royal Society (URF/R/180008, RGF/EA/181062).

The authors acknowledge the use of resources provided by the Isambard-AI National AI Research Resource (AIRR). Isambard-AI is operated by the University of Bristol and is funded by the UK Government Department for Science, Innovation and Technology (DSIT) via UK Research and Innovation; and the Science and Technology Facilities Council (ST/AIRR/I-A-I/1023).

\printbibliography
\end{document}

%% file: sections/1.introduction.tex
\clearpage
\section{Introduction}
\label{sec:intro}

Transformers~\cite{VaswaniAttention} have become central to modern machine learning due to their flexible attention mechanisms, which model long-range interactions across diverse data modalities. Query-based transformer decoders, in particular, have become influential in set-to-set prediction tasks such as object detection and image segmentation~\cite{detr,mask2former}, instance segmentation of 3D point clouds~\cite{Schult2023Mask3D}, and emerging scientific applications including particle reconstruction~\cite{van2023vertex,trackml_maskformer,GLOW}. These architectures use a learned set of object queries that interact with an encoded representation of the input through cross-attention, allowing the model to infer a variable number of underlying entities in an end-to-end fashion.

\begin{figure*}[h]
\centering
\begin{subfigure}[t]{0.48\textwidth}
\centering
\includegraphics[width=\linewidth]{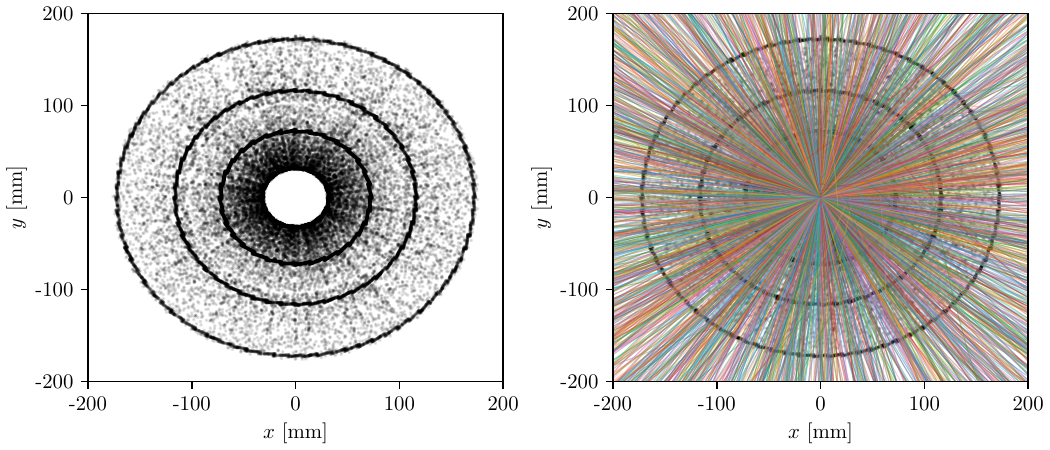}
\caption{
Transverse $x$--$y$ projection. Left: detector hits. Right: corresponding charged-particle trajectories.
}
\label{fig:event_display_xy}
\end{subfigure}
\hfill
\begin{subfigure}[t]{0.48\textwidth}
\centering
\includegraphics[width=\linewidth]{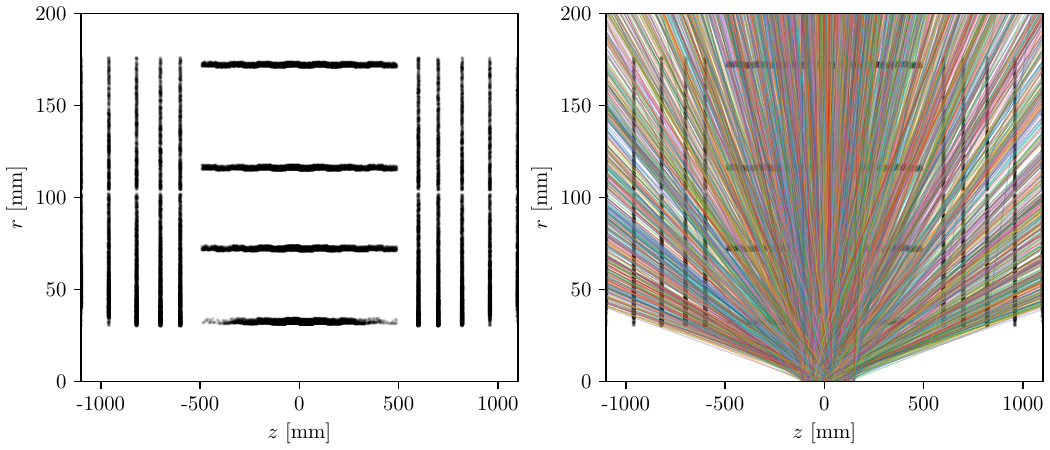}
\caption{
Longitudinal $r$--$z$ projection, where $z$ is the beam direction and $r$ is the radial distance from the beamline. Left: detector hits. Right: corresponding trajectories.
}
\label{fig:event_display_rz}
\end{subfigure}
\caption{
Event-display illustration of charged-particle tracking in two detector projections. The reconstruction task is to partition sparse detector measurements, or hits, into sets corresponding to individual particle trajectories. This sparse, irregular, and high-multiplicity structure contrasts with the dense lattice structure of images and motivates the query-based set-prediction formulation studied in this work.
}
\label{fig:event_display_intro}
\end{figure*}

Query-based transformer architectures have been widely applied to dense, regularly gridded modalities such as images and video, where spatial tokens lie on a fixed lattice and can be exploited efficiently. In contrast, particle-physics detectors produce sparse and irregularly structured measurements from events with large and variable object multiplicities. We study this setting in the context of charged-particle tracking, building on Ref.~\cite{trackml_maskformer}, where thousands of particle trajectories must be reconstructed from tens of thousands of detector measurements, or \textit{hits}, in each collision event. Figure~\ref{fig:event_display_intro} illustrates this reconstruction problem in two detector projections, showing the sparse hit measurements and the corresponding charged-particle trajectories. The model takes hits as input and represents particle trajectories using decoder queries that predict query--hit association masks.

In this setting, decoder cross-attention becomes a dominant cost, scaling as $\mathcal{O}(N_q \times N_h)$ with the number of object queries $N_q$ and input tokens $N_h$. For charged-particle tracking, $N_h$ is the number of input hits retained for trajectory reconstruction, while $N_q$ is the number of candidate trajectory queries processed by the decoder. While efficient self-attention mechanisms such as sliding-window attention~\cite{liu2021swin} and FlashAttention~\cite{2022arXiv220514135D} mitigate encoder cost, the decoder remains a bottleneck because the baseline masked cross-attention implementation materialises the full query--hit attention matrix.

A further source of inefficiency is the use of an input-independent set of learned parametric query embeddings for data whose object content is strongly event dependent. This produces two distinct limitations. First, the decoder must infer event-specific trajectory structure from generic query representations, even though the encoder outputs already contain rich geometric information at the level of individual detector hits. Second, the number of reconstructable trajectories varies substantially across collision events. A fixed query budget must therefore be chosen conservatively to cover the largest expected multiplicities. Consequently, many queries correspond to no physical trajectory in typical events while still incurring the full cost of query--hit attention.

Compared with the fixed-query MaskFormer tracking model of Ref.~\cite{trackml_maskformer}, this work changes the decoder formulation in two central ways. First, it replaces the fixed learned query bank with event-conditioned queries generated from predicted first-hit encoder embeddings, allowing both query content and query multiplicity to adapt to each event. Second, it exploits the induced azimuthal ordering of these dynamic queries to replace learned, irregular masked cross-attention with a geometry-informed local sparse attention pattern. These changes alter the decoder interface, reduce the dominant query--hit attention cost, and enable operation in higher-occupancy and heterogeneous detector regimes not demonstrated in Ref.~\cite{trackml_maskformer}.

The first component of the proposed architecture is \emph{dynamic query generation} (DQ), in which decoder queries are initialised directly from encoder-level hit representations on a per-event basis. This makes query cardinality event-dependent, allowing decoder computation to scale with the trajectory multiplicity present in each event rather than a conservative worst-case budget. It also grounds the query embeddings in event-specific geometric information, reducing the burden on the decoder to construct object representations from generic learned vectors.

In cylindrical tracking detectors, the azimuthal coordinate $\phi$ provides a physically meaningful ordering of hits, with hits from the same trajectory typically occupying a narrow neighbourhood in this ordered representation. By initialising decoder queries from selected $\phi$-ordered hit embeddings, dynamic query generation induces an aligned ordering over the query set. As a result, the relevant query--hit interactions concentrate near a band-diagonal structure.

We exploit this induced structure through \emph{Local Strided Cross-Attention} (LSCA), which replaces dense decoder cross-attention with a stride-aligned local attention support. LSCA restricts each query to attend to a fixed neighbourhood around its corresponding diagonal region in the $\phi$-ordered hit sequence, avoiding geometrically implausible query--hit interactions and exposing structured sparsity for block-sparse attention kernels.

We evaluate the resulting performance trade-off using two standard tracking metrics. Tracking efficiency is defined as the fraction of reconstructable particles matched to a predicted trajectory, while the fake rate is defined as the fraction of predicted trajectories that do not correspond to any reconstructable particle. Together, these metrics quantify complementary failure modes associated with missed physical trajectories and spurious reconstructed tracks.

The contributions of this work are threefold. First, we introduce a dynamic query generation mechanism that replaces fixed, learned decoder queries with input-conditioned queries derived directly from encoder features. This event-conditioned query construction adapts both query cardinality and query content to the observed event, aligning decoder capacity with the number and structure of candidate trajectories. As a result, it reduces the burden on the computationally expensive decoder and enables a lighter decoder architecture while improving reconstruction quality relative to the fixed-query baseline.

Second, we exploit the ordering inherited by the dynamic queries from the $\phi$-ordered hit embeddings used for their initialisation. This query ordering aligns candidate trajectories with their corresponding regions of the hit sequence, causing the dominant query--hit associations to concentrate near the diagonal of the cross-attention map. We use this geometry-induced structure to reformulate decoder cross-attention as an explicitly local operation, implemented with strided diagonal attention masks that expose structured sparsity for sparse attention kernels.

Third, we present a quantitative evaluation of tracking performance, memory consumption, and inference latency across multiple detector configurations and decoder variants. The proposed architecture reduces decoder memory usage and inference time while extending the model to higher-occupancy and more challenging regimes, including lower transverse-momentum thresholds, wider kinematic acceptance, and heterogeneous pixel--strip detector inputs.

%% file: sections/2.related.tex
\section{Related Work}
\label{sec:related}

Transformers have been widely adopted for structured prediction tasks due to their flexible, content-based interactions, yet their quadratic complexity remains a challenge for high-resolution or high-multiplicity inputs. Prior work has approached this limitation from several complementary directions.

\vspace{-0.5em}
\paragraph{Query-based transformer decoders.} Object detection in vision has increasingly shifted toward query-based transformer architectures, originating with DETR~\cite{detr}, which formulates detection as a set-prediction task using a fixed set of learned object queries that attend globally to image features. In their standard form, these architectures use dense decoder cross-attention over an $N_q\times N_h$ interaction space, which limits scalability for high-resolution or high-multiplicity data such as those encountered in particle physics detectors. Although subsequent variants modify the attention mechanism or the queries themselves to improve convergence or reduce computational overhead (e.g.\ using deformable sampling, token filtering, anchor- or box-parameterised queries, or denoising objectives), the spatial priors incorporated in these vision models (e.g.\ anchor grids or learned reference points) assume Cartesian image geometry and do not directly transfer to irregular sensors.

\vspace{-0.5em}
\paragraph{Local Attention in Transformers.}
Vision and language models have incorporated locality to reduce self-attention cost. Swin Transformer~\cite{liu2021swin} uses non-overlapping windows with shifted windowing across layers, while Longformer~\cite{beltagy2020longformer} and BigBird~\cite{zaheer2020bigbird} employ sparse attention patterns combining local windows, dilated global attention, and random sparsity. 
Fast Segment Anything~\cite{Zhao2023FastSA} demonstrated that imposing structured locality can substantially accelerate mask-based segmentation models without degrading accuracy. 
These designs, however, operate on dense image-like inputs and primarily modify self-attention, whereas charged-particle tracking involves sparse, irregular point clouds where a bottleneck lies in the \emph{decoder} cross-attention.

\vspace{-0.5em}
\paragraph{Efficient Sparse Attention Implementations.}
Recent attention kernels such as FlashAttention~\cite{2022arXiv220514135D} and FlexAttention~\cite{flexAttn} provide large practical speedups by reducing memory movement and avoiding unnecessary computation. FlashAttention avoids materialising the dense attention matrix by using fused kernels that tile the attention computation and minimise transfers to and from GPU memory. FlexAttention extends this style of optimisation to user-defined attention variants by allowing custom score and mask modifications to be lowered to optimised fused kernels. For sparse masks, FlexAttention constructs a block-sparse \texttt{BlockMask} representation of the attention pattern, in which the score matrix is divided into fixed-size tiles and fully masked tiles are skipped during attention computation. This avoids materialising and computing over the full $N_q \times N_h$ attention matrix, allowing memory and compute to scale approximately with the number of active attention blocks rather than the full interaction space. Our LSCA masks are explicitly designed to expose this structured block sparsity.

\vspace{-0.5em}
\paragraph{Scientific Applications.}
Query-based transformer decoders have recently been explored in high-energy physics for sparse object reconstruction. Our earlier MaskFormer-based tracking model~\cite{trackml_maskformer} demonstrated that learned object queries can recover charged-particle trajectories with high efficiency ($94\%$) and low fake rate ($0.7\%$), with an inference time of \SI{99}{ms} on an NVIDIA A100 for trajectories of particles with transverse momentum $\pT>1$~GeV in the TrackML~\cite{trackmlA,trackmlB} dataset.

Locality-sensitive hashing (LSH) has been used to reduce encoder complexity in scientific point-cloud data. HEPT~\cite{miao2024hept} introduced an LSH-bucketed encoder with near linear scaling, and HEPTv2~\cite{govil2024hept} extended this to charged-particle tracking using a decoder architecture closely related to our earlier MaskFormer-style formulation~\cite{trackml_maskformer}. HEPTv2 achieves fast inference (\SI{28}{ms}) with high efficiency (99\%) but retains dense cross-attention, resulting in dense $N_q\times N_h$ attention maps and fake rates of 11\% on the full TrackML detector.

Graph neural networks~\cite{gnn_original_paper} (GNNs) remain the most mature machine-learning solution deployed for collider tracking. The GNN4ITk pipeline achieves 95--98\% efficiency with fake rates as low as $10^{-5}$ for $p_\textrm{T}>1$~GeV in simulation of the future ATLAS Inner Tracker, and its GPU-accelerated implementation reaches 37--95~ms latency on an NVIDIA H100, depending on model size and precision~\cite{ATL-PHYS-PUB-2025-046}.

%% file: sections/3.HEP.tex
\section{Charged-Particle Tracking and Detector Geometry}
\label{sec:hep}

At the CERN Large Hadron Collider (LHC)~\cite{Evans:2008zzb}, bunches of approximately $10^{11}$ protons circulate in opposite directions around a 27~km ring and collide at the centres of multi-purpose detectors such as ATLAS~\cite{PERF-2007-01} and CMS~\cite{CMS-CMS-00-001}. Bunch crossings occur every 25~ns, corresponding to a rate of 40~MHz. Under high-luminosity operating conditions, a single bunch crossing, hereafter referred to as an \emph{event}, contains on average 200 simultaneous proton--proton collisions. Each collision occurs at a spatial production point, referred to as a vertex. A single event therefore produces a complex final state containing many charged particles, with a typical event containing $\mathcal{O}(10^3)$ charged particles whose passage through the silicon tracking detector generates $\mathcal{O}(10^4)$ spatial measurements, or \emph{hits}. Classical algorithms for reconstructing the underlying charged-particle trajectories from these sparse, irregularly distributed measurements scale poorly with hit multiplicity, making efficient trajectory reconstruction one of the central algorithmic challenges in modern collider physics.

The silicon tracking detector is composed of concentric cylindrical layers of sensors around the beam axis ($z$-axis), complemented by endcap disks in the forward regions. The innermost layers are pixel detectors, which provide fine-grained two-dimensional position measurements with high spatial precision. At larger radii, the detector transitions to strip sensors, which offer coarser spatial measurements and enable efficient coverage of much larger areas. In the baseline reconstruction setting studied in Ref.~\cite{trackml_maskformer}, only pixel hits were considered. In this work, we extend the problem to include strip measurements.

We define the target trajectory population in terms of \emph{reconstructable particles}, defined as charged particles that lie within the kinematic and geometric acceptance and leave sufficient detector measurements to support reconstruction. In the pixel-only settings, a particle is considered reconstructable if it leaves at least three pixel hits. In the strip-inclusive setting, where both pixel and strip measurements are used, a particle must leave at least seven hits in total, including at least three pixel hits. The full configuration-dependent selections are specified in Section~\ref{sec:experiments}.

\begin{figure*}[h]
    \centering
    \begin{subfigure}[t]{0.44\textwidth}
        \centering
        \includegraphics[width=\linewidth]{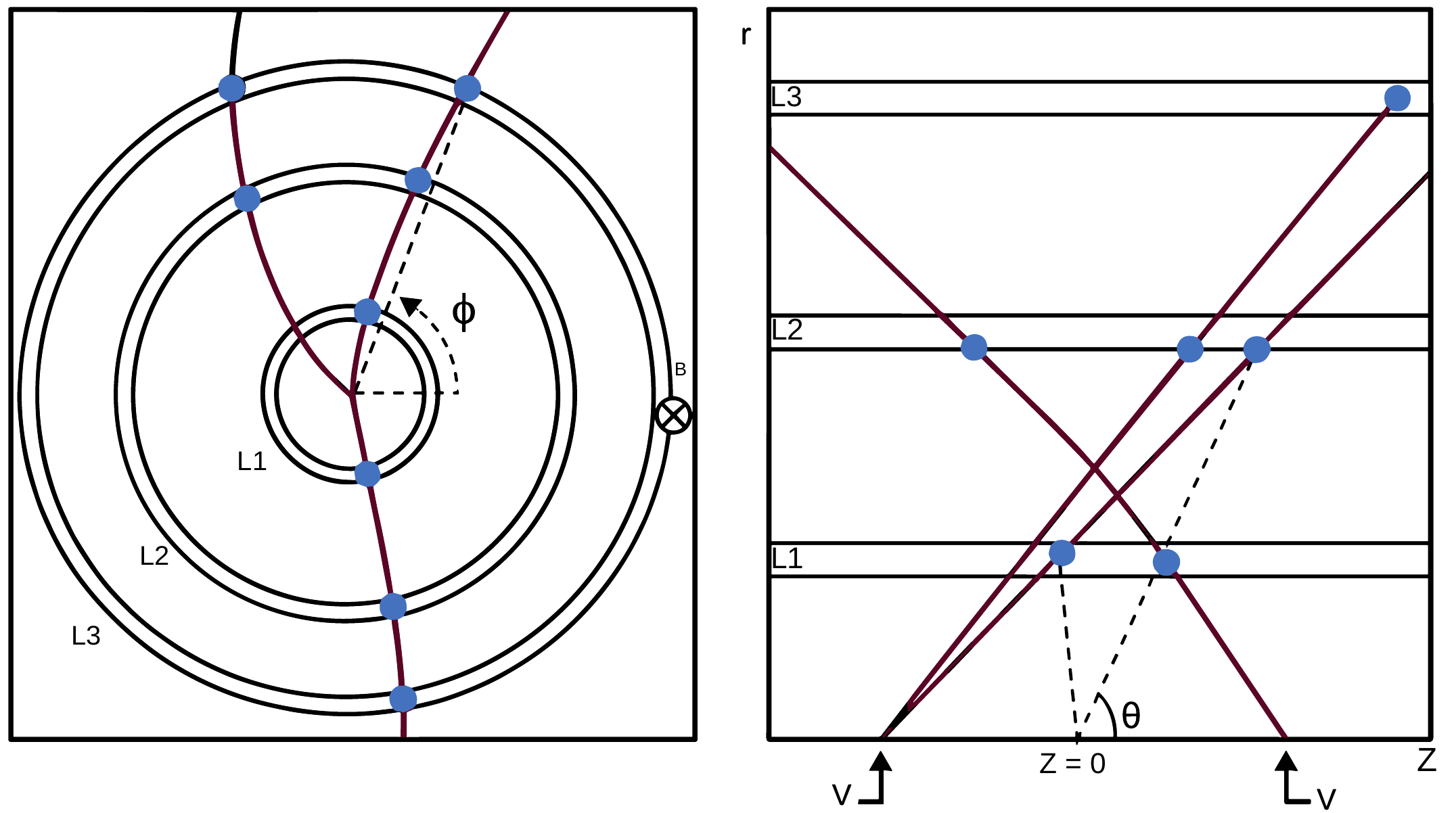}
        \caption{
        Schematic comparison of charged-particle locality in the transverse $(r,\phi)$ and longitudinal $(r,z)$ detector projections relative to the LHC beamline.
        }
        \label{fig:phi_eta_geometry_combined}
    \end{subfigure}
    \hfill
    \begin{subfigure}[t]{0.52\textwidth}
        \centering
        \includegraphics[width=\linewidth]{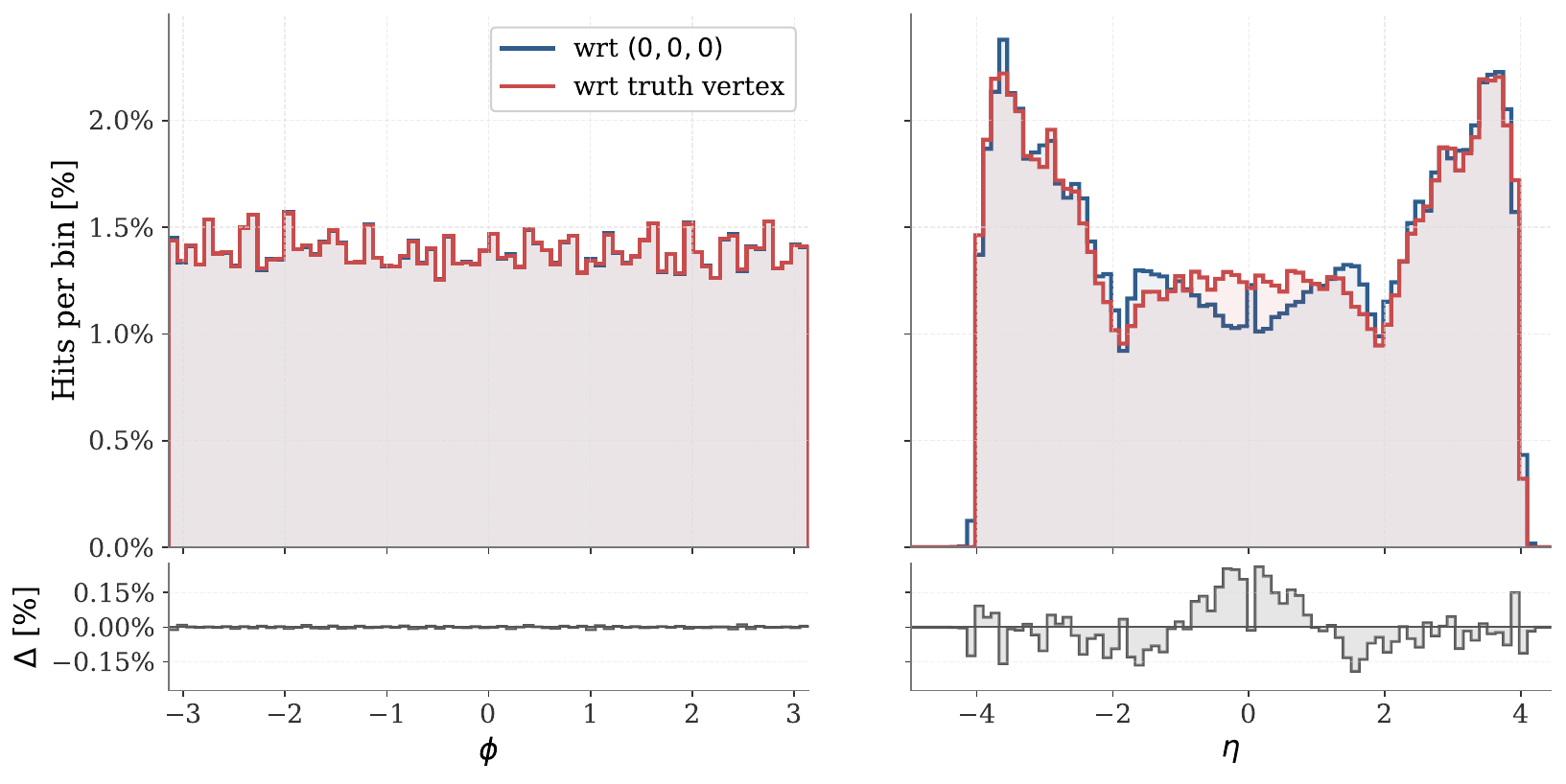}
        \caption{
        Dataset-level hit occupancy in $\phi$ and $\eta$ under different vertex assumptions.
        }
        \label{fig:phi_eta_hist_validation}
    \end{subfigure}
    \caption{
    Comparison of locality in $\phi$ and $\eta$ for charged-particle hits. Here $z$ denotes the beam axis, $r$ the radial coordinate, and $\phi$ the azimuthal angle; L1--L3 denote successive detector layers, $B$ the solenoidal magnetic-field direction, and $v$ the longitudinal production vertex. Panel~(a) illustrates that charged-particle trajectories remain locally coherent in azimuth, while the polar angle inferred in the $(r,z)$ projection depends on the production vertex. Panel~(b) shows that the hit distribution in $\phi$ is approximately stable under different vertex assumptions, whereas the inferred $\eta$ distribution changes substantially when computed relative to the detector origin or the true production vertex.
    }
    \label{fig:phi_vs_eta_locality}
\end{figure*}

The tracking detector is immersed in a strong solenoidal magnetic field approximately aligned with the beam axis, conventionally denoted as the $z$ axis. The transverse plane is the plane perpendicular to this beam axis, described by the radial coordinate $r$ and azimuthal angle $\phi$. Under the Lorentz force, charged particles follow approximately helical trajectories, with approximately circular motion in the transverse $(r,\phi)$ plane with curvature inversely proportional to transverse momentum $\pT$, and approximately linear evolution in $z$, as depicted in Figure~\ref{fig:phi_vs_eta_locality}.

Charged-particle tracking is commonly decomposed into two stages, a combinatorial pattern-recognition stage that associates detector hits into candidate trajectories, followed by a fitting stage that estimates the trajectory parameters. The present work focuses on the first stage, which is typically the more challenging component because it requires solving a discrete assignment problem over many ambiguous hit combinations before the trajectory parameters are known.

This naturally casts tracking as a sparse object-reconstruction problem over an irregular detector geometry expressed in cylindrical coordinates $(r,\phi,z)$. A key geometric property in the mid- to high-\pT regime considered here is that transverse bending between successive detector layers remains modest. As a result, hits belonging to the same charged particle remain confined to a narrow and smoothly varying band in azimuth $\phi$, as illustrated in Fig.~\ref{fig:phi_vs_eta_locality}. The particle density is also approximately uniform in $\phi$, so sorting hits by azimuth induces a stable and physically meaningful notion of locality.

In contrast, the polar angle $\theta$, or equivalently the pseudorapidity $\eta=-\ln\tan(\theta/2)$, does not provide an equally robust locality variable. Although $\eta$ is conserved along an ideal trajectory when defined relative to the true production point, its inference from detector hits depends on the unknown longitudinal origin of the particle. In addition, the particle density varies with $\eta$, making fixed-size local windows in this coordinate substantially less uniform across events. As shown in Fig.~\ref{fig:phi_vs_eta_locality}, geometric locality is therefore most naturally and robustly expressed in $\phi$.

In an idealised detector, each charged particle would generate one precise measurement on each traversed layer, all lying on a helical trajectory emerging from the interaction region. Real detectors are more complex. The magnetic field is non-uniform, interactions with detector material induce multiple scattering, in which particles undergo small stochastic deflections as they traverse matter, energy loss perturbs the trajectory curvature, detector inefficiencies lead to missing measurements, and electronic noise introduces spurious hits. These effects make tracking a large-scale structured set-partitioning problem in which hits must be assigned to their generating particles despite ambiguity, missing information, and detector-specific resolution effects.

Event occupancy varies significantly across events, but the number of reconstructable particles is strongly correlated with the total number of associated hits. Since particles typically leave a comparable number of measurements as they traverse the detector, the particle multiplicity grows approximately linearly with hit multiplicity. This provides a direct link between the observed input size, the underlying combinatorial complexity of the reconstruction task, and the number of queries needed for a particular event.

The layered detector geometry also provides natural representative measurements for each reconstructable trajectory. In this work, we define the first hit of a trajectory as its innermost associated detector measurement, and use this hit as the representative measurement for that trajectory. This choice provides the basis for the dynamic query construction introduced in Section~\ref{sec:dynamic_queries}, in which predicted first-hit candidates are used to initialise the decoder queries. Their ordering in the $\phi$-sorted input sequence then defines the corresponding query order.

%% file: sections/4.model.tex
\section{Model}
\label{sec:model}

Our architecture builds upon the query-based tracking model introduced in Ref.~\cite{trackml_maskformer}, which formulates charged-particle reconstruction as a global query--hit assignment problem. Following a MaskFormer-style formulation, charged-particle tracks are represented by a set of object queries, with each query corresponding to a candidate trajectory.

\begin{figure}[t]
\centering
\includegraphics[width=0.8\textwidth]{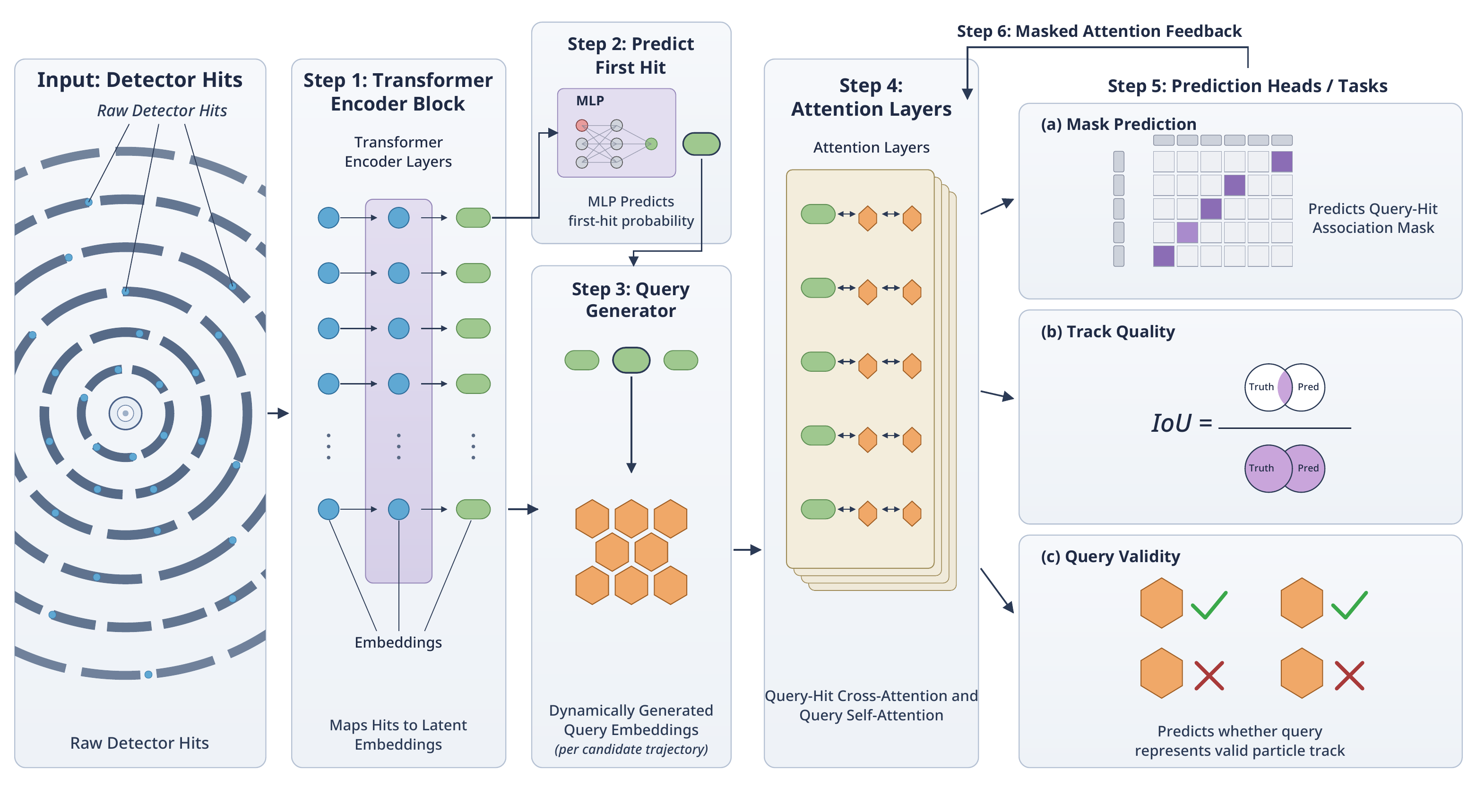}
\caption{
Overview of the dynamic-query tracking architecture.
Raw detector-hit coordinates are encoded by a Transformer encoder, and an auxiliary first-hit classifier identifies candidate track seeds.
The selected first-hit embeddings initialise an event-dependent set of decoder object queries aligned with physical track candidates.
A Transformer decoder refines these queries through query self-attention and query--hit cross-attention, producing query--hit assignment masks.
Task-specific heads predict query--hit associations, query validity, and an auxiliary track-quality score.
In the masked-attention decoder variant, intermediate query--hit masks are also fed back into subsequent decoder layers to gate cross-attention.
}
\label{fig:transformer_tracking_model}
\end{figure}

The model retains the encoder--decoder Transformer structure of Ref.~\cite{trackml_maskformer}. A Transformer encoder maps the $\phi$-ordered detector hits to latent hit embeddings using windowed self-attention in azimuth. Restricting self-attention to local regions of size $w$ exploits the approximate locality of charged-particle trajectories in $\phi$ and reduces the encoder complexity from $\mathcal{O}(N_h^2)$ to $\mathcal{O}(N_h \times w)$, yielding linear scaling in the number of hits $N_h$ for fixed $w$. An overview of the resulting architecture, including dynamic query generation, decoder refinement, and the auxiliary prediction heads, is shown in Fig.~\ref{fig:transformer_tracking_model}.

The principal architectural change introduced in this work is the replacement of fixed, generic, learned decoder queries with dynamically generated queries constructed from encoder-level hit embeddings. Dynamic query generation is described in Section~\ref{sec:dynamic_queries}.

Aside from query initialisation, the decoder follows the same iterative refinement principle as Ref.~\cite{trackml_maskformer}. In each decoder layer, query self-attention coordinates candidate tracks, cross-attention aggregates information from detector-hit embeddings, and a query--hit assignment head predicts soft association masks from the current query and hit embeddings. In the masked-attention variant, these predicted masks gate subsequent cross-attention through a masked attention operator, restricting each query's receptive field to its predicted hit subset. Section~\ref{sec:diagonal_structure} shows how, under dynamic query initialisation, the geometry inherited by the query embeddings induces a band-diagonal structure in these masks, allowing masked cross-attention to be reformulated as a local windowed operation with substantially reduced memory cost.

Alongside the query--hit assignment head, the final refined query embeddings are processed by two auxiliary query-level prediction heads. A \textit{query-validity} head predicts whether a given query corresponds to a physical trajectory. We additionally introduce a \textit{track-quality} head, which is not present in Ref.~\cite{trackml_maskformer}. This head regresses a per-track quality score defined as the intersection-over-union (IoU) between the predicted and ground-truth hit masks.

As in Ref.~\cite{trackml_maskformer}, a substantial fraction of detector hits originate from low-\pT particles or detector noise and do not contribute to the target trajectory population considered in this study. To reduce the computational burden on the tracking model, a dedicated hit-filtering network is applied upstream of track reconstruction. We employ the same hit-filter architecture as in the previous work.

\subsection{Dynamic Query Generation}
\label{sec:dynamic_queries}

The number of reconstructable trajectories varies substantially from event to event, making fixed learned decoder query sets inherently mismatched to individual samples. A fixed query set must either under-provision complex events or waste capacity on simpler ones. Dynamic query generation replaces the fixed learned query set with an event-dependent set of query embeddings derived from the encoder-level hit representations.

For each reconstructable trajectory, we define the first hit as the innermost associated detector measurement. First-hit candidates therefore provide an unambiguous and physically motivated representative hit for each track.\footnote{The use of the first hit is an operational design choice rather than a unique requirement of the dynamic-query framework. Other representative measurements, such as the last hit in the ordered pixel detector sequence, could also be used to initialise event-dependent queries.} Using the corresponding encoder embeddings to initialise the decoder queries yields an event-dependent query set that is approximately one-to-one with the reconstructable trajectory population and aligned with physical track candidates.

Representative hits are identified by an auxiliary classifier applied to the encoder outputs. For each hit embedding, a lightweight task head consisting of a dense layer of dimension $256$ followed by a sigmoid activation predicts the probability that the hit is a first hit, defined operationally as the innermost hit associated with a charged particle. Hits whose probabilities exceed a model-specific validation-tuned threshold are selected as candidate first hits. Since the encoder outputs preserve the sequence order of the $\phi$-sorted input hits, the selected candidates retain this $\phi$ ordering. Their encoder embeddings are then used directly to initialise the decoder queries, so the resulting dynamic query sequence is $\phi$-ordered by construction.

To stabilise training and ensure bounded computational cost, the number of initialised queries per event is capped at a fixed maximum $N_\text{max}$. If more than $N_\text{max}$ hits exceed the first-hit classification threshold, only the $N_\text{max}$ hits with the highest predicted probabilities are retained. The parameter $N_\text{max}$ therefore defines the maximum number of trajectories that can be reconstructed in a single event, and is chosen to cover the maximum track multiplicity across all training events.

\begin{figure}[h]
    \centering
    \includegraphics[width=0.4\linewidth]{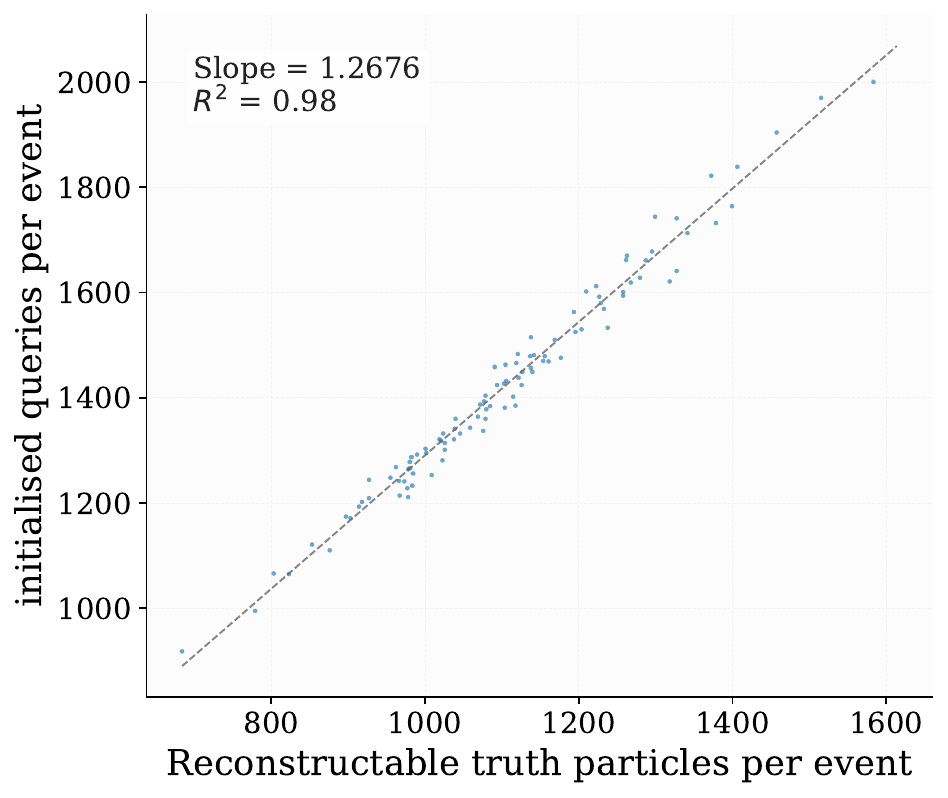}
    \caption{
    Relationship between the number of initialised dynamic queries and reconstructable truth particles per event. A strong linear correlation ($R^2 = 0.98$) is observed, indicating that query cardinality adapts closely to event complexity.
    }
    \label{fig:dq_scaling}
\end{figure}

Figure~\ref{fig:dq_scaling} and Table~\ref{tab:dq_stats} compare the number of selected first-hit candidates with the number of reconstructable particles in each event. The close correlation between these quantities shows that the first-hit classifier adapts query multiplicity to event occupancy, while the systematic surplus of selected candidates reflects the validation-selected operating point. The first-hit threshold is tuned to optimise final tracking performance rather than to match the reconstructable-particle multiplicity exactly. This favours high seed efficiency as redundant local track hypotheses, initialised by false-positive first-hit candidates, can be rejected by the query-validity head, whereas missing the first hit of a true trajectory removes its seed query and makes subsequent recovery difficult. The resulting modest over-provisioning therefore provides a safety margin against under-seeding.

\begin{table}[h]
    \centering
    \begin{tabular}{lcc}
        \toprule
        & mean $\pm$ std & max \\
        \midrule
        Dynamic queries & $1400 \pm 200$ & $2000$ \\
        Reconstructable truth particles & $1100 \pm 160$ & $1700$ \\
        Surplus dynamic queries & $300 \pm 50$ & $460$ \\
        \bottomrule
    \end{tabular}
    \caption{
    Summary statistics for dynamic-query counts and reconstructable ground-truth particles per event in the nominal pixel-only configuration with $p_T \geq 1~\mathrm{GeV}$, $|\eta| \leq 4$, and at least three pixel hits. Dynamic queries are initialised from predicted innermost-hit candidates. The final row reports the per-event surplus relative to the number of reconstructable particles.
    }
    \label{tab:dq_stats}
\end{table}

The resulting dynamically generated queries are propagated through the decoder using an iterative refinement principle similar to that of the baseline architecture, with query self-attention and query--hit cross-attention used to refine track representations and predict hit assignments. Since queries are initialised from first-hit encoder embeddings, they are already aligned with candidate physical tracks. This reduces the permutation degeneracy of the output queries as, rather than requiring the decoder to determine which generic query should represent which trajectory, each query is anchored to a specific detector hit and hence to a local track hypothesis. The decoder can therefore focus on refining query--hit associations rather than constructing trajectory representations from generic event-agnostic queries. This enables more efficient use of decoder capacity, with attention concentrated on an informative event-dependent query set.

This event-specific query initialisation enables a substantially lighter decoder than that used in Ref.~\cite{trackml_maskformer}. We use $3$ decoder layers instead of $8$ and remove bidirectional query--hit cross-attention. Section~\ref{sec:results} shows that these simplifications reduce memory usage and inference latency while retaining competitive tracking performance.

\subsection{Local Strided Cross-Attention (LSCA)}

Dynamic query initialisation induces an ordered query set whose alignment with the $\phi$-sorted detector hits gives rise to structured, azimuthally local query--hit interactions. We exploit this property to reformulate decoder cross-attention as a local banded operation.

\subsubsection{Diagonal Structure in Query--Hit Attention}
\label{sec:diagonal_structure}

As described in Section~\ref{sec:dynamic_queries}, dynamic query generation selects first-hit candidates from the $\phi$-ordered hit sequence and uses their encoder embeddings to initialise the decoder queries. The query sequence therefore inherits the same $\phi$ ordering as the hit sequence. Since hits from the same trajectory remain local when the hit sequence is sorted by $\phi$, the hits relevant to a given query are expected to appear near the corresponding diagonal region of the ordered query--hit interaction map.

\begin{figure}[h]
    \centering
    \begin{subfigure}[t]{0.3\linewidth}
        \centering
        \includegraphics[width=\linewidth]{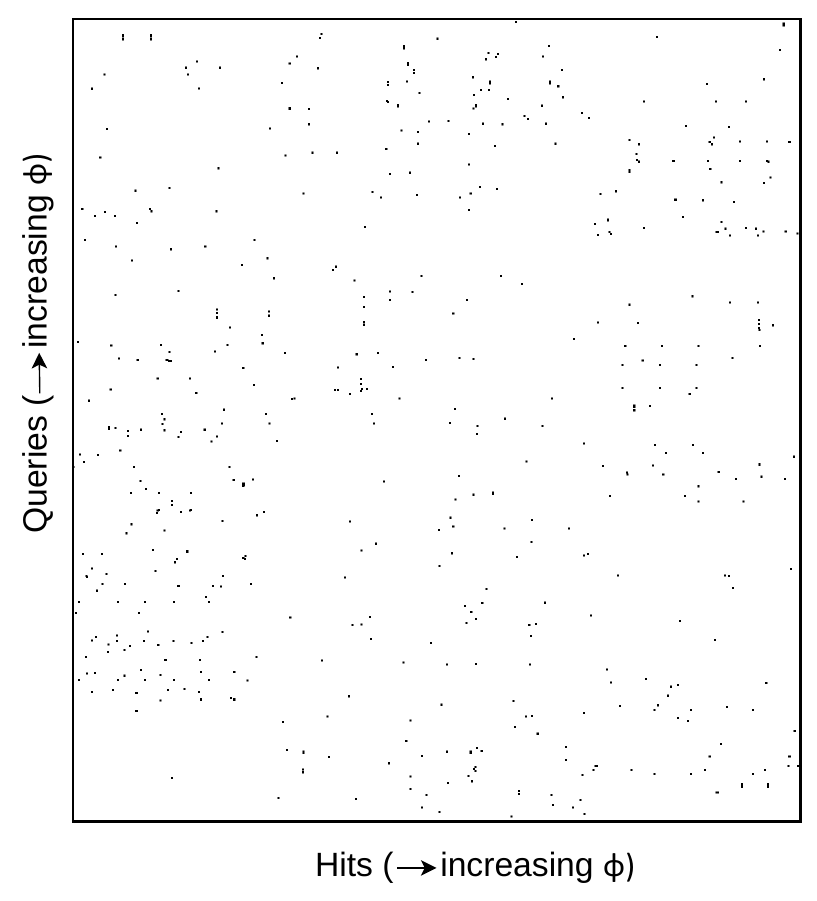}
        \caption{Fixed learned queries}
        \label{fig:diagonal_attention_static}
    \end{subfigure}
    \hspace{0.04\linewidth}
    \begin{subfigure}[t]{0.305\linewidth}
        \centering
        \includegraphics[width=\linewidth]{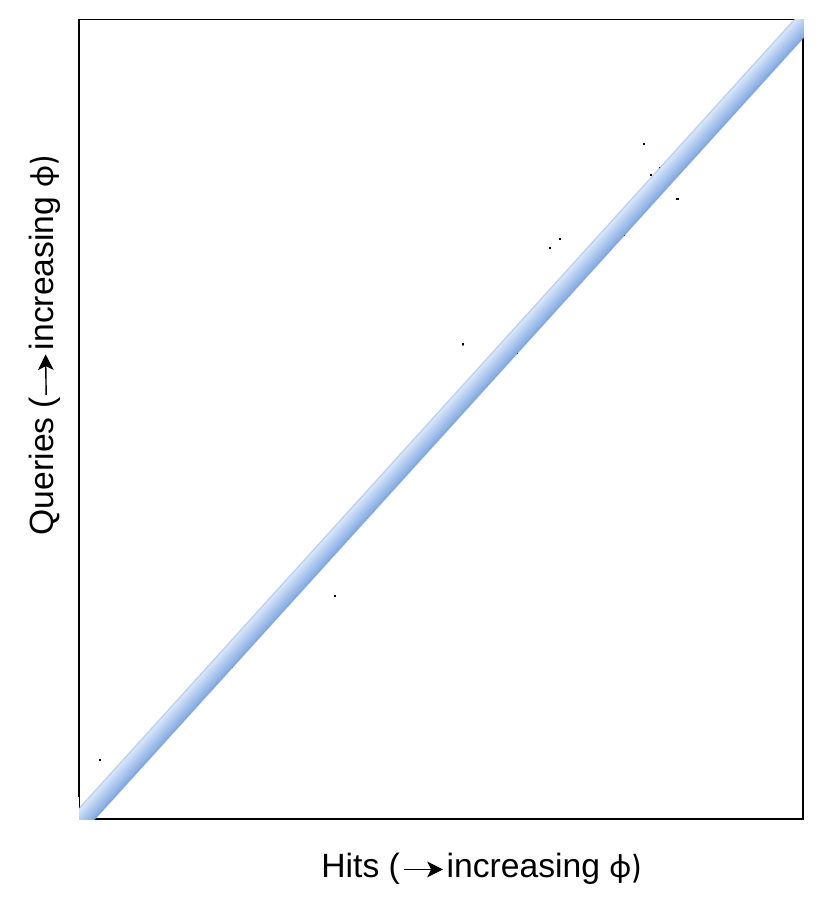}
        \caption{Dynamic queries ($\phi$-ordered)}
        \label{fig:diagonal_attention_dynamic}
    \end{subfigure}
    \caption{
    Visualisation of predicted query--hit assignment masks for (a) fixed learned and (b) dynamic query initialisation. Fixed learned queries produce diffuse masks with no clear geometric alignment to the $\phi$-ordered hits. Dynamic queries initialised from $\phi$-ordered hit embeddings produce a pronounced banded structure. The blue shaded band illustrates the locality region used by LSCA.
    }
    \label{fig:diagonal_attention}
\end{figure}

This geometric alignment is reflected in the attention-support masks used by the masked-attention decoder. In the MaskFormer-style decoder, the mask used to gate a cross-attention layer is predicted from the current query embeddings and the decoder hit embeddings. Hits belonging to the same physical trajectory are expected to have similar latent representations, so the interaction between a query embedding and the hit embeddings provides an early estimate of promising query--hit associations. Dynamic query initialisation strengthens this effect because each query embedding is initialised directly from a representative hit embedding on the candidate trajectory. Since both the seed hits and the full hit sequence are ordered by $\phi$, the predicted association masks are expected to concentrate near the diagonal of the query--hit map. This expected band-diagonal structure is visible in the intermediate query--hit assignment masks used to gate subsequent cross-attention layers, as shown in Fig.~\ref{fig:diagonal_attention}. As a result, much of the dense query--hit interaction space corresponds to geometrically implausible associations, motivating an explicitly local cross-attention mechanism.

\subsubsection{Local Strided Cross-Attention with FlexAttention}
\label{sec:lca_flex}

Under dynamic query initialisation, the predicted query--hit association masks that gate subsequent decoder cross-attention concentrate near the diagonal. This motivates reformulating cross-attention as a local operation. LSCA restricts each query to attend only within a fixed-width window around its corresponding diagonal position in the $\phi$-ordered hit sequence, as illustrated in Fig.~\ref{fig:diagonal_attention_dynamic}.

Conceptually, this departs from the original MaskFormer paradigm, in which attention masks are learned as latent proposals defining which inputs each query should attend to. Instead, LSCA specifies the cross-attention support using a geometry-based prior arising from the consistent ordering of dynamic queries and detector hits, together with the azimuthal locality of charged-particle trajectories. The set of hits available to each query is therefore determined by the query index in the ordered dynamic-query sequence, rather than by a learned mask prediction.

Let $i \in \{0,\ldots,N_q-1\}$ index the $\phi$-ordered dynamic queries and let
$j \in \{0,\ldots,N_h-1\}$ index the $\phi$-ordered hits. Since the hit sequence
is longer than the query sequence, $N_h > N_q$, we define a stride
\begin{equation}
s = \frac{N_h}{N_q},
\end{equation}
which maps query indices to approximate positions on the hit axis. The centre of the local hit window associated with query $i$ is then
\begin{equation}
c_i = \left(i+\frac{1}{2}\right)s .
\end{equation}
For an LSCA window of width $W$, we define a binary attention-support mask
\begin{equation}
A_{ij}^{\mathrm{LSCA}} =
\begin{cases}
1, & \left| j - c_i \right| \leq \frac{W}{2},\\
0, & \mathrm{otherwise}.
\end{cases}
\label{eq:lsca_support}
\end{equation}
Entries with $A_{ij}^{\mathrm{LSCA}}=0$ are excluded from cross-attention. In implementation, $c_i$ is rounded to the nearest integer hit index when constructing the mask. The periodicity of $\phi$ is handled by wrapping the ordered hit sequence at the $-\pi/\pi$ boundary, so that hits close in azimuth remain adjacent even when they lie on opposite ends of the sorted sequence. Thus, each query attends to a fixed-width band of hits centred on its stride-mapped diagonal position in the query--hit attention matrix.

We implement this attention-support pattern using PyTorch FlexAttention~\cite{flexAttn}, which represents the restriction as a block-sparse \texttt{BlockMask}. Fully masked blocks outside the diagonal band are skipped before score computation and are never materialised, so memory and compute scale with the number of active diagonal blocks rather than the full $N_q \times N_h$ interaction space. For a fixed local window size, and an approximately fixed query-to-hit ratio, query--hit cross-attention therefore scales approximately linearly with the number of hits, in contrast to the $\mathcal{O}(N_q N_h)$ scaling of cross-attention implementations that materialise the full query--hit interaction space.

%% file: sections/5.experiments.tex
\section{Experiments}
\label{sec:experiments}

\subsection{Training Setup}
\label{sec:training_setup}

The encoder maps each input detector hit to a latent representation of dimension $d=256$ using a linear embedding of the per-hit input features, following the representation used in Ref.~\cite{trackml_maskformer}. These features include the three-dimensional hit position and additional detector-level information associated with the measurement. Geometric information is incorporated through positional encodings defined in cylindrical coordinates, with a cyclic representation of the azimuthal angle $\phi$ to respect its periodicity. The resulting hit embeddings are processed by a stack of Transformer encoder layers with model dimension $d$ and feed-forward dimension $2d$, using windowed self-attention over $\phi$-ordered hits. For the pixel-only configurations, the self-attention window size is set to $w=512$, chosen by coarse hyperparameter optimisation. Configuration-dependent encoder and decoder depths are given in Table~\ref{tab:model_hyperparameters}.

As described in Section~\ref{sec:dynamic_queries}, decoder queries are generated dynamically from the encoder outputs. These dynamically generated object queries are processed by three decoder layers. Increasing the number of decoder layers improves reconstruction performance but incurs additional inference cost; in the pixel-only configurations, a depth of three layers is selected as a trade-off between these factors. Both encoder and decoder Transformer blocks employ a HybridNorm architecture~\cite{zhuo2025hybridnorm}. Initial embedding networks use SwiGLU activations \cite{shazeer2020glu}, while task-specific heads employ SiLU activations \cite{elfwing2018sigmoid}, a choice found to improve performance.

Query--hit assignment scores are obtained by first projecting each refined query embedding to a mask token, and then computing the dot product between this mask token and the decoder hit embeddings, followed by a sigmoid activation. The resulting soft masks are binarised to determine whether a given hit is assigned to a given track candidate. In parallel, a query-validity head predicts whether each object query corresponds to a physical track using a binary classifier with a single hidden layer of size $128$, suppressing spurious queries by masking their outputs. A separate track-quality head predicts the intersection-over-union between the predicted and target query--hit association masks, providing a learned estimate of reconstruction quality that can be used to control the purity of the selected track candidates.

The training loss is the weighted sum of the loss terms from each of the tasks.
\begin{equation}
    L =
    0.1 L_\text{Valid}+ 
    \underbrace{   
 2 L_\text{Dice} + 25 L_\text{Focal}}_{L_{\text{Mask}}} +
     L_{\text{Quality}} + L_{\text{DQ}}.
\end{equation}

For both the query-validity prediction, $L_{\text{Valid}}$, and the first hit prediction, $L_{\text{DQ}}$, a binary cross-entropy loss is used. The mask loss $L_{\text{Mask}}$ is a combination of a Dice loss $L_{\text{Dice}}$~\cite{dice} and a focal loss $L_{\text{Focal}}$~\cite{2017arXiv170802002L}. The track-quality loss $L_{\mathrm{Quality}}$ is the mean squared error between the predicted track-quality score and the true intersection-over-union of the predicted and target hit masks.

The intermediate outputs from each decoder layer are used to compute auxiliary loss terms which are included in the total loss. The query-validity supervision in the first decoder layer is disabled to allow query embeddings to specialise before classification. The track-quality loss is not included in auxiliary loss terms.

Loss weights are selected to provide a balanced trade-off between the different task objectives. The loss is defined using the optimal bipartite matching between the predicted and target objects, as computed with an efficient linear assignment problem solver \cite{10.1145/3442348}. This ensures that the loss is invariant over permutations of the object queries \cite{trackml_maskformer}.

Optimisation is performed using the Lion optimiser with a one-cycle learning rate schedule, with initial, peak, and final learning rates of $1\times10^{-5}$, $5\times10^{-5}$, and $1\times10^{-5}$ respectively, and weight decay $1\times10^{-5}$. Each training epoch consists of 8500 events, and models are trained for $30$ epochs.

\subsection{Model Variants}
\label{sec:model_variants}

Target particles, i.e. reconstructable trajectories, are defined by
\begin{equation}
\pT \geq \pT^{\min}, \qquad
|\eta| \leq \eta_{\max}, \qquad
N_{\mathrm{hits}} \geq N_{\min}.
\end{equation}
For pixel-only configurations, reconstructability requires at least $N_{\min}^{\mathrm{pixel}}=3$ pixel hits. For strip-inclusive configurations, we additionally require at least $N_{\min}^{\mathrm{total}}=7$ total hits. These criteria define the trajectory population used for supervision and evaluation in each detector configuration.

We evaluate three detector configurations of increasing reconstruction complexity:
\begin{itemize}
\item \textbf{Pix1.0}: $\pT \geq 1~\mathrm{GeV}$, $|\eta| \leq 4$, pixel detector only;
\item \textbf{Pix0.6}: $\pT \geq 0.6~\mathrm{GeV}$, $|\eta| \leq 4$, pixel detector only;
\item \textbf{PixStrip1.0}: $\pT \geq 1~\mathrm{GeV}$, $|\eta| \leq 4$, pixel and strip detectors.
\end{itemize}

The Pix1.0 and Pix0.6 configurations isolate the pixel-detector reconstruction problem while varying the transverse-momentum threshold. Lowering $\pT^{\min}$ increases the target-particle multiplicity and introduces more strongly curved trajectories, thereby increasing the combinatorial ambiguity of the hit-association task. These two configurations therefore provide a controlled setting in which to study reconstruction performance and computational scaling as the effective event complexity increases. The efficiency improvements introduced in this work enable full pixel-detector acceptance up to $|\eta| \leq 4$ at $\pT \geq 0.6~\mathrm{GeV}$, extending beyond the reduced-acceptance low-\pT setting used in the previous fixed-query formulation.

The PixStrip1.0 configuration extends the reconstruction problem by incorporating strip measurements in addition to pixel hits. This is not a like-for-like scaling comparison with the pixel-only configurations, since the strip-inclusive task uses a stricter reconstructability definition, requires hit association over a heterogeneous detector input, and uses additional architectural components.

Pixel and strip hits are first processed by separate input embedding networks, reflecting their different readout geometries and position resolutions, and are then concatenated before the shared encoder. Dynamic queries are initialised exclusively from pixel-hit embeddings, and decoder query refinement uses cross-attention with the pixel-hit representations only. This design treats the high-resolution pixel measurements as a learned track-seeding representation, with the resulting query embeddings defining candidate trajectories. These shared query embeddings are then used to predict associations to both pixel and strip hits. Instead of predicting a single association mask over the union of pixel and strip measurements, the model uses separate association heads for the two detector types, producing detector-type-specific masks from the same query representation.

Many alternative mechanisms for incorporating strip information are possible, including direct pixel--strip decoder cross-attention or more tightly coupled pixel--strip association heads. The configuration studied here should therefore be interpreted as an exploratory extension of the dynamic-query and LSCA framework to a more complete detector geometry, rather than as an optimised strip-inclusive architecture. Its timing and memory measurements should accordingly be interpreted as the cost of this strip-inclusive reconstruction configuration, rather than as pure scaling measurements with respect to retained hit multiplicity.

Configuration-dependent settings are chosen separately for each benchmark configuration. The first-hit classification threshold and LSCA window size are tuned on the validation set, while the maximum number of dynamic queries, $N_{\mathrm{DQ}}^{\max}$, is set to provide sufficient capacity for the observed track multiplicities. Encoder and decoder depths are selected to balance reconstruction performance against latency and memory cost. The LSCA window controls the locality of query--hit cross-attention and is selected to balance hit-association efficiency and purity. The resulting settings are summarised in Table~\ref{tab:model_hyperparameters}.

\begin{table}[h!]
\centering
\caption{
Configuration-dependent settings for the three detector configurations.
Here $N_{\mathrm{DQ}}^{\max}$ denotes the maximum number of dynamic queries retained per event.
The Pix1.0 and~Pix0.6 configurations use the shared pixel-only architecture, while the PixStrip1.0 configuration uses the strip-inclusive architecture with pixel-only decoder cross-attention and detector-specific association heads.
}
\label{tab:model_hyperparameters}
\begin{tabular}{lccccc}
\toprule
Model & $N_{\mathrm{DQ}}^{\max}$ & First-hit threshold & LSCA window & Encoder layers & Decoder layers \\
\midrule
Pix1.0      & 2000 & 0.1 & 64  & 8  & 3 \\
Pix0.6      & 3900 & 0.3 & 64  & 8  & 3 \\
PixStrip1.0 & 1800 & 0.3 & 256 & 15 & 6 \\
\bottomrule
\end{tabular}
\end{table}

\subsection{Experimental Setup and Metrics}\label{sec:experimental_setup}

For each detector configuration, we train and evaluate two decoder variants, Masked Attention (MA) and Local Strided Cross-Attention (LSCA). Within a fixed detector configuration, all architectural components except the decoder cross-attention are held fixed, ensuring that performance differences between MA and LSCA can be attributed to the attention formulation.

Computational performance is measured by per-event inference latency and peak allocated GPU memory during both training and inference. Memory values correspond to peak allocated CUDA memory. Experiments are performed on a single NVIDIA A100 GPU using mixed precision (bfloat16). Track reconstruction quality is assessed by matching reconstructed trajectories to reconstructable truth particles. Following Ref.~\cite{trackml_maskformer}, we use the double-majority (DM) criterion, which requires majority overlap between a reconstructed track and its truth particle in both directions. The stricter perfect-matching criterion requires that all hits from a truth particle are assigned to a single reconstructed track, with no additional hits, and is used as an auxiliary diagnostic where relevant.

Tracking efficiency measures the fraction of reconstructable particles successfully matched to a reconstructed track. The \textit{double-majority efficiency} $\epsilon_{\mathrm{DM}}$ is defined as the fraction of reconstructable truth particles with at least one DM-matched reconstructed track, while the \textit{perfect-match efficiency} $\epsilon_{\mathrm{perf}}$ is the fraction with at least one perfectly matched reconstructed track.

The \textit{fake rate} $f_{\mathrm{DM}}$ is the fraction of reconstructed tracks not assigned to any truth particle under the DM matching criterion. Since each truth particle is allowed to match to at most one reconstructed track, duplicate reconstructions of an already matched particle are counted as unmatched reconstructed tracks and therefore contribute to the fake rate. Thus, $f_{\mathrm{DM}}$ quantifies the fraction of reconstructed tracks corresponding to spurious hit combinations, incomplete assignments, and duplicate reconstructions of already matched particles.

\subsection{Hit Filtering and Effective Event Complexity}

As introduced in Section~\ref{sec:model}, a learned hit-filtering model is applied upstream of track reconstruction to remove detector hits unlikely to originate from reconstructable particles. The filter threshold sets the trade-off between noise rejection and particle-level retention and is tuned independently for each benchmark configuration. After filtering, the effective reconstruction scale is set by the retained hit multiplicity rather than the raw detector occupancy. The resulting event occupancies and target multiplicities are summarised in Table~\ref{tab:event_complexity}.

\begin{table*}[h!]
\centering
\small
\setlength{\tabcolsep}{4pt}
\begin{tabular}{lccccc}
\toprule
Model
& Pre-filter hits (k)
& Post-filter hits (k)
& Reco. particle retention (\%)
& Reco. particles/event
& Max reco. particles \\
\midrule
Pix1.0 & 56.7 & 9.7 & 98.6 & $1100 \pm 160$ & 1700 \\
Pix0.6 & 56.7 & 20.3 & 99.1 & $2600 \pm 340$ & 3900 \\
PixStrip1.0 & 94.5 & 18.2 & 98.6 & $1000 \pm 150$ & 1600 \\
\bottomrule
\end{tabular}
\caption{
Configuration-dependent event complexity.
Columns report the mean hit multiplicity before and after filtering, the fraction of reconstructable particles retained after filtering, and the mean, standard deviation, and maximum number of reconstructable truth particles per event.
}
\label{tab:event_complexity}
\end{table*}

%% file: sections/6.results.tex
\section{Results}
\label{sec:results}

\subsection{Tracking Performance under Local Cross-Attention}
\label{sec:model_a_tracking}

\begin{figure}[h!]
    \centering
    \begin{subfigure}[h!]{0.48\textwidth}
        \centering
        \includegraphics[width=\linewidth]{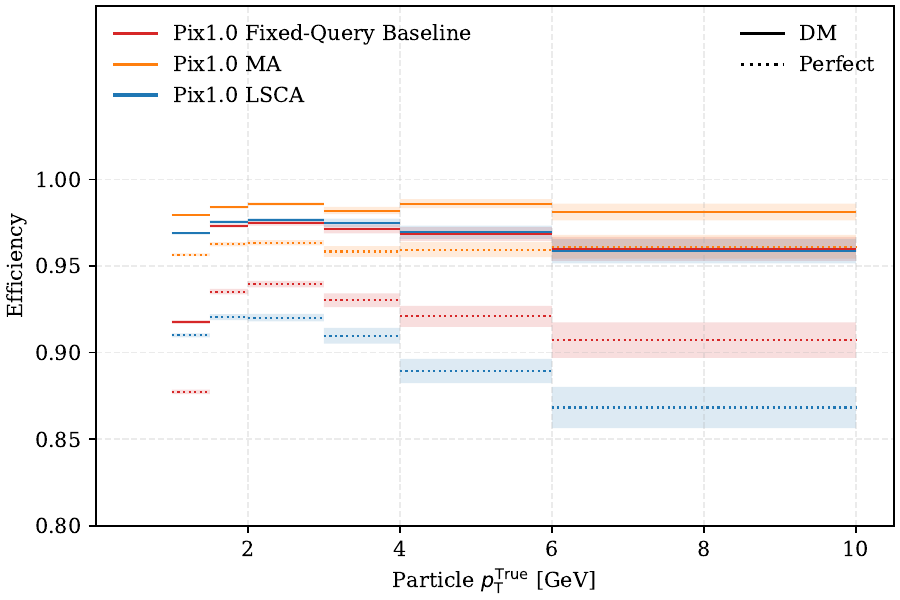}
        \caption{Efficiency as a function of transverse momentum \pT.}
        \label{fig:model_a_efficiency_pt}
    \end{subfigure}
    \hfill
    \begin{subfigure}[h!]{0.48\textwidth}
        \centering
        \includegraphics[width=\linewidth]{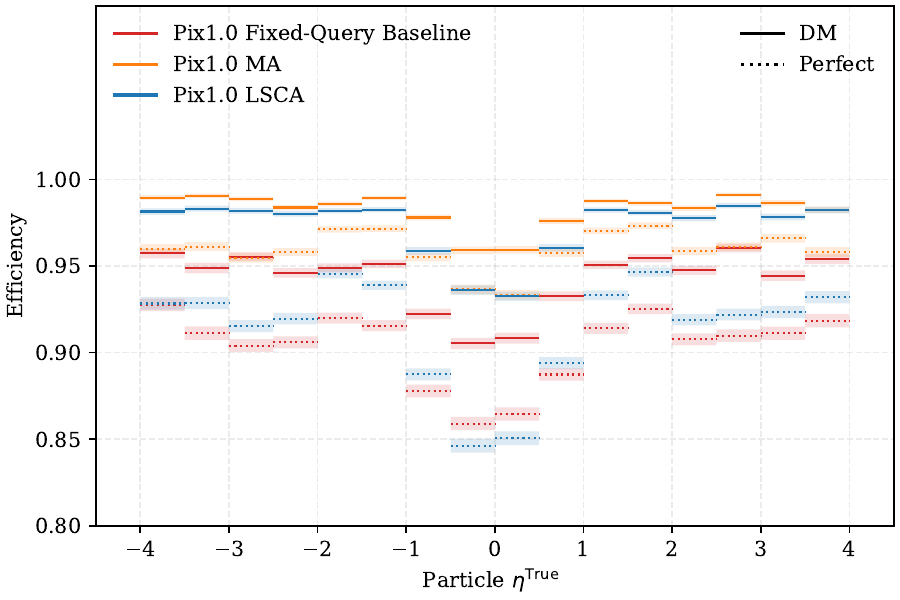}
        \caption{Efficiency as a function of pseudorapidity $\eta$.}
        \label{fig:model_a_efficiency_eta}
    \end{subfigure}
    \caption{
    Track reconstruction efficiency for the Pix1.0 configuration as a function of transverse momentum and pseudorapidity.
    Both double-majority and perfect-match efficiencies are shown for the fixed-query baseline, the DQ+MA model, and the DQ+LSCA decoder.
    }
    \label{fig:model_a_efficiency}
\end{figure}

We first evaluate track reconstruction in the nominal pixel-only configuration, Pix1.0, comparing the dynamic-query Masked Attention (DQ+MA) and Local Strided Cross-Attention (DQ+LSCA) models with the fixed-query Masked Attention (FQ+MA) baseline of Ref.~\cite{trackml_maskformer}. Figure~\ref{fig:model_a_efficiency} shows the corresponding double-majority and perfect-match tracking efficiencies as functions of transverse momentum and pseudorapidity, while aggregate metrics for the pixel-only configurations are summarised in Table~\ref{tab:pixel_tracking_performance}.

\begin{table}[h]
\centering
\caption{
Tracking and filtering performance in the pixel-only configurations.
Reported metrics are the fake rate $f_{\mathrm{DM}}$, the combined hit-filter plus tracking double-majority efficiency $\epsilon_{\mathrm{DM}}^{\mathrm{filter+track}}$, and the combined hit-filter plus tracking perfect-match efficiency $\epsilon_{\mathrm{perfect}}^{\mathrm{filter+track}}$.
}
\label{tab:pixel_tracking_performance}
\begin{tabular}{llccc}
\toprule
Model & Decoder & $f_{\mathrm{DM}}$ [\%] & $\epsilon_{\mathrm{DM}}^{\mathrm{filter+track}}$ [\%] & $\epsilon_{\mathrm{perfect}}^{\mathrm{filter+track}}$ [\%] \\
\midrule
Pix1.0 & Fixed Query Baseline & 0.7 & 94.1 & 90.4 \\
& DQ+MA                & 0.3 & 98.1 & 95.8 \\
& DQ+LSCA              & 1.0 & 97.1 & 91.2 \\
\midrule
Pix0.6 & DQ+MA                & 0.3 & 98.0 & 93.2 \\
& DQ+LSCA              & 0.4 & 97.6 & 91.5 \\
\bottomrule
\end{tabular}
\end{table}

In the Pix1.0 configuration, the DQ+MA model improves the combined filter-plus-tracking double-majority efficiency to over $98\%$, a gain of $4.0$ percentage points relative to the fixed-query baseline, while reducing the fake rate by almost $60\%$. The perfect-match efficiency also increases by $5.4$ percentage points to nearly $96\%$, indicating that the dynamic-query architecture improves not only track recovery under the double-majority criterion, but also the fidelity of the full hit assignment. This demonstrates that the dynamic-query architecture, in combination with the associated architectural modifications, improves reconstruction quality while enabling a lighter decoder.

Replacing DQ+MA with DQ+LSCA introduces the expected accuracy--efficiency trade-off. In the Pix1.0 configuration, the combined filter-plus-tracking double-majority efficiency decreases by $1.0$ percentage point and the perfect-match efficiency by $4.6$ percentage points, while the fake rate increases from $0.3\%$ to $1.0\%$. This degradation is consistent with the reduced flexibility of LSCA relative to masked attention, since LSCA uses a fixed geometry-based attention window rather than a learned query-specific mask. Nevertheless, DQ+LSCA remains competitive with the previous fixed-query MA baseline, improving the combined filter-plus-tracking double-majority efficiency by $3.0$ percentage points and the perfect-match efficiency by $0.8$ percentage points, although with a higher fake rate. As shown in Section~\ref{sec:pixel_scaling}, this competitive reconstruction performance is obtained with substantially reduced latency and memory usage.

The reduced decoder cost of the dynamic-query architecture enables extension to the higher-occupancy Pix0.6 configuration, in which the transverse-momentum threshold is lowered from $1~\mathrm{GeV}$ to $0.6~\mathrm{GeV}$. As shown in Table~\ref{tab:event_complexity}, relative to the Pix1.0 configuration, this more than doubles both the number of reconstructable particles and the retained hit multiplicity, producing a substantially more challenging query--hit association problem.

\begin{figure}[h!]
\centering
\begin{subfigure}[h!]{0.46\textwidth}
\centering
\includegraphics[width=\linewidth]{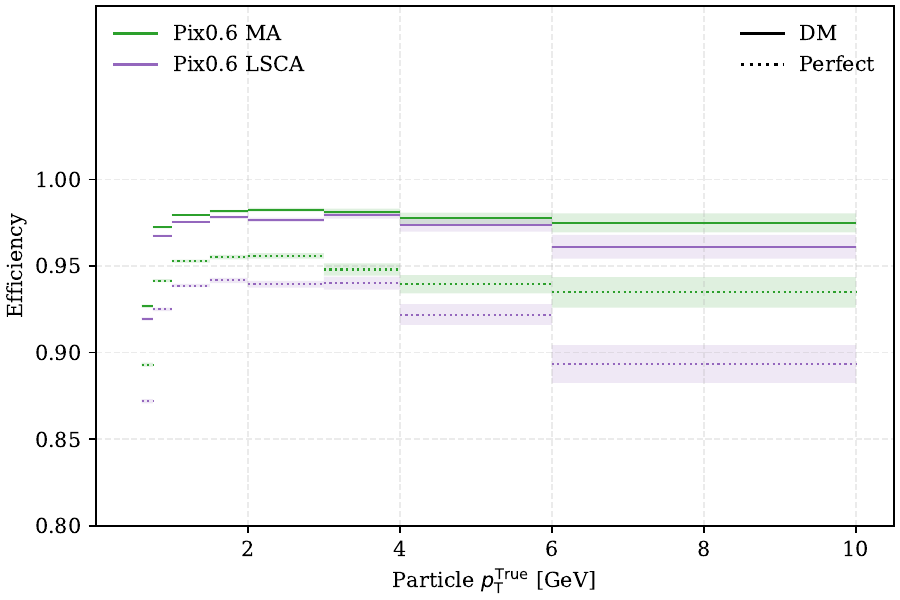}
\caption{Efficiency as a function of true particle transverse momentum $p_\mathrm{T}^{\mathrm{true}}$.}
\label{fig:pixel_extension_efficiency_pt}
\end{subfigure}
\hfill
\begin{subfigure}[h!]{0.48\textwidth}
\centering
\includegraphics[width=\linewidth]{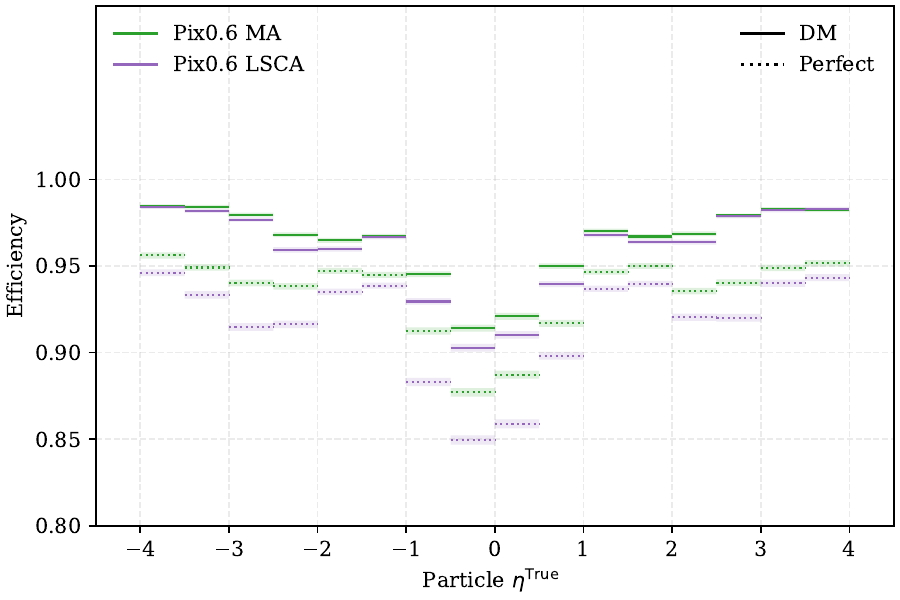}
\caption{Efficiency as a function of true particle pseudorapidity $\eta^{\mathrm{true}}$.}
\label{fig:pixel_extension_efficiency_eta}
\end{subfigure}
\caption{
Tracking efficiency for the high-occupancy pixel-only configuration, Pix0.6.
Solid curves show double-majority efficiency, while dotted curves show perfect-match efficiency, for the DQ+MA and DQ+LSCA decoder variants.
}
\label{fig:model_b_efficiency}
\end{figure}

Tracking performance in this high-occupancy pixel regime is summarised in the Pix0.6 rows of Table~\ref{tab:pixel_tracking_performance}, with the corresponding kinematic dependence shown in Fig.~\ref{fig:model_b_efficiency}. Despite the increased reconstruction complexity, both decoder variants retain high end-to-end efficiency. Replacing masked attention with LSCA again introduces only a modest accuracy--efficiency trade-off. Relative to DQ+MA, the combined filter-plus-tracking double-majority efficiency decreases by only $0.4$ percentage points, while the stricter perfect-match efficiency decreases by $1.7$ percentage points and the fake rate increases from $0.3\%$ to $0.4\%$.

These results represent a significant extension of the operating regime relative to the previous fixed-query formulation. Despite the substantially larger target multiplicity and retained hit set in the Pix0.6 configuration, both decoder variants maintain high reconstruction efficiency over the studied $|\eta| \leq 4$ pixel-detector acceptance. This demonstrates that the approach can scale to the lower-momentum, higher-occupancy conditions required for more realistic tracking applications.

\subsection{Runtime and Memory Scaling in Pixel-Only Regimes}
\label{sec:pixel_scaling}

We next examine the computational scaling of the two dynamic-query decoder variants. Using the pixel-only configurations introduced above, we compare memory usage and per-event inference latency as functions of retained hit multiplicity and dynamic query count. The Pix1.0 configuration probes the nominal occupancy regime, while the Pix0.6 configuration tests the higher-occupancy setting induced by the lower transverse-momentum threshold.

As summarised in Table~\ref{tab:systems_all}, both proposed dynamic-query architectures reduce latency relative to the previous FQ+MA baseline. The fixed-query baseline timings are taken from the corresponding baseline measurement, while the dynamic-query end-to-end timings combine the updated tracking measurements with the independently measured hit-filter latency.

In the Pix1.0 configuration regime, the fixed-query baseline requires $76.0 \pm 9.0~\mathrm{ms}$ for tracking-only inference, while DQ+MA reduces this to $48.0 \pm 2.9~\mathrm{ms}$, corresponding to a reduction of approximately $37\%$. DQ+LSCA reduces tracking-only latency further to $37.1 \pm 14.0~\mathrm{ms}$, corresponding to an overall reduction of approximately $51\%$ relative to the fixed-query baseline. The benefit of LSCA becomes more pronounced in the higher-occupancy Pix0.6 configuration, where tracking-only latency decreases from $71.9 \pm 19.8~\mathrm{ms}$ for DQ+MA to $34.2 \pm 8.6~\mathrm{ms}$ for DQ+LSCA, a $52\%$ relative reduction.

\begin{table*}[h]
\centering
\caption{
Computational performance for the pixel-only configurations. `Trk Time' denotes tracking-only inference latency, `HF + Trk Time' includes the upstream hit filter, and `Train Mem.' and `Infer. Mem.' denote peak allocated GPU memory during training and inference, respectively. Timing values are reported as the mean per event, with quoted spreads corresponding to the RMS over events. Memory values correspond to peak allocated GPU memory.
}
\label{tab:systems_all}
\begin{tabular}{llcccc}
\toprule
Model & Decoder
& Trk Time [ms]
& HF + Trk Time [ms]
& Train Mem. [GB]
& Infer. Mem. [GB] \\
\midrule
Pix1.0 & FQ+MA & $76.0 \pm 9.0$ & $99.0 \pm 11.0$ & 23.8 & 2.2 \\
       & DQ+MA       & $48.0 \pm 2.9$ & $63.7 \pm 6.6$ & 3.9 & 0.9 \\
       & DQ+LSCA     & $37.1 \pm 14.0$ & $52.8 \pm 15.2$ & 3.0 & 0.2 \\
\midrule
Pix0.6 & DQ+MA       & $71.9 \pm 19.8$ & $87.4 \pm 20.3$ & 11.1 & 3.8 \\
       & DQ+LSCA     & $34.2 \pm 8.6$ & $49.7 \pm 9.8$ & 9.8 & 0.6 \\
\bottomrule
\end{tabular}
\end{table*}

\begin{figure*}[h!]
    \centering
    \begin{subfigure}[h!]{0.48\textwidth}
        \centering
        \includegraphics[width=\linewidth]{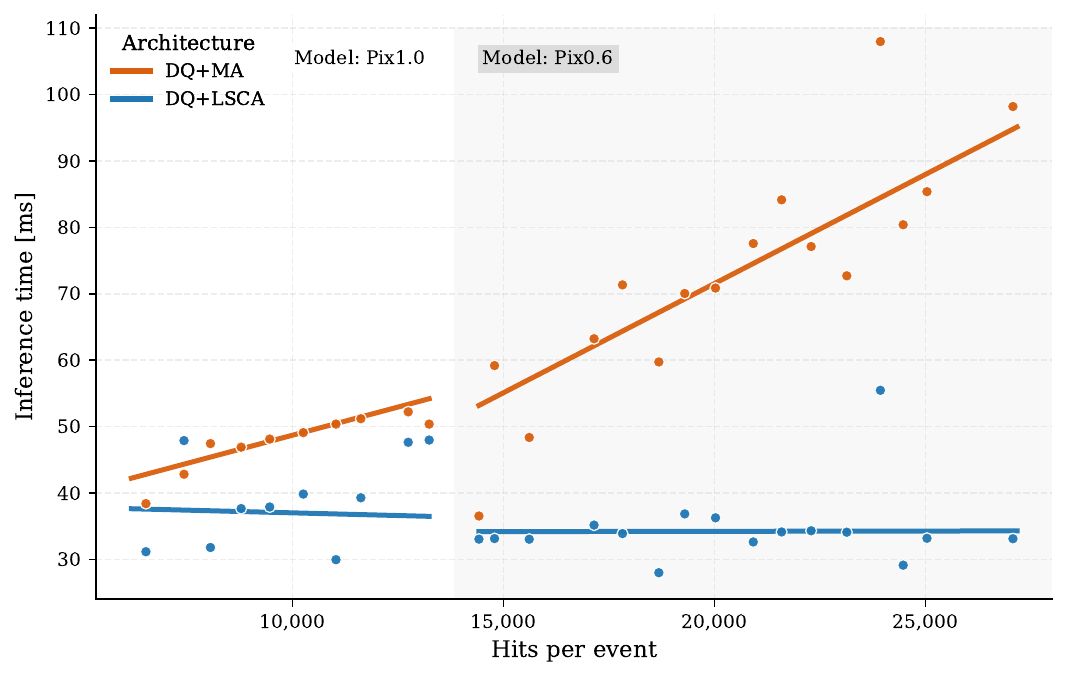}
        \caption{Binned per-event inference latency as a function of the number of retained hits.}
        \label{fig:regime_runtime_scaling_hits}
    \end{subfigure}
    \hfill
    \begin{subfigure}[h!]{0.48\textwidth}
        \centering
        \includegraphics[width=\linewidth]{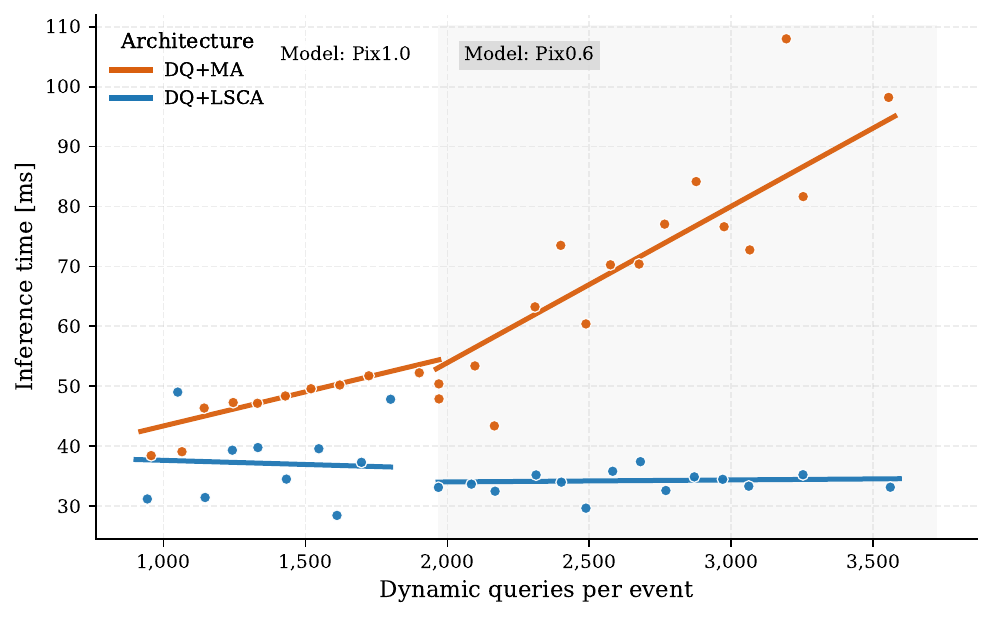}
        \caption{Binned per-event inference latency as a function of dynamic query count.}
        \label{fig:regime_runtime_scaling_queries}
    \end{subfigure}
    \caption{
    Tracking-only per-event inference latency across pixel-only detector configurations of increasing complexity, shown as a function of retained hit multiplicity and dynamic query count.
    Colours distinguish the decoder variants, Masked Attention (DQ+MA) and Local Strided Cross-Attention (DQ+LSCA).
    Markers show mean latency values within bins of the corresponding event-level quantity, with error bars indicating the standard deviation within each bin.
    The fitted trends discussed in the text are obtained from the unbinned event-level measurements.
    The unshaded region indicates the Pix1.0 configuration regime, while the shaded region indicates the Pix0.6 configuration.
    }
    \label{fig:regime_runtime_scaling}
\end{figure*}

Figure~\ref{fig:regime_runtime_scaling} shows the empirical tracking-only latency scaling of the two decoder variants after binning events by retained hit multiplicity and dynamic query count. DQ+MA latency increases with both event-level quantities, whereas DQ+LSCA remains approximately flat over the same range. Linear fits to the unbinned measurements confirm this trend. As a function of retained hit multiplicity, the fitted DQ+MA slopes are $1.7\times10^{-3}~\mathrm{ms/hit}$ and $3.3\times10^{-3}~\mathrm{ms/hit}$ for Models~Pix1.0 and~Pix0.6, respectively, while the corresponding DQ+LSCA slopes are consistent with zero or negligible. The same pattern is observed as a function of dynamic query count. DQ+MA has fitted slopes of $1.1\times10^{-2}~\mathrm{ms/query}$ and $2.6\times10^{-2}~\mathrm{ms/query}$, whereas DQ+LSCA remains approximately flat. These results indicate that replacing learned masked attention with a local diagonal attention pattern, coupled to a block-sparse implementation, largely removes the empirical dependence of decoder latency on both retained hit multiplicity and query multiplicity over the studied event range. The near-flat behaviour does not imply that LSCA has no query- or hit-dependent cost. Rather, the cross-attention component is sufficiently reduced that its incremental cost is subdominant to other timing overheads in this regime.

\begin{figure*}[h!]
    \centering
    \begin{subfigure}[h!]{0.48\textwidth}
        \centering
        \includegraphics[width=\linewidth]{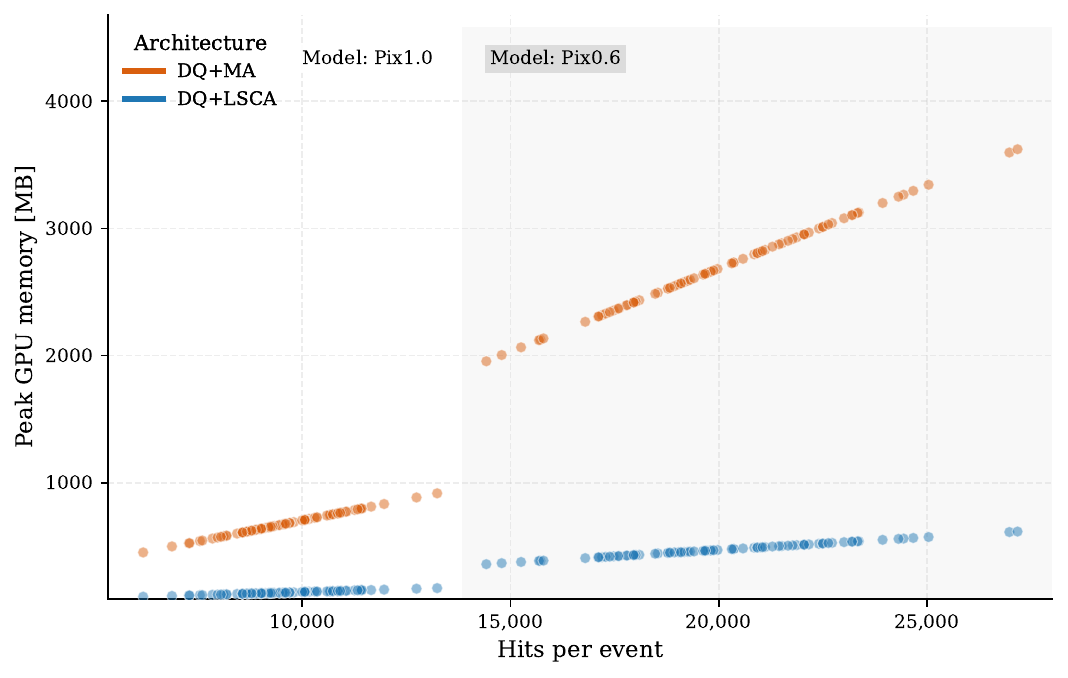}
        \caption{Peak inference GPU memory usage as a function of retained hit multiplicity.}
        \label{fig:regime_memory_scaling_hits}
    \end{subfigure}
    \hfill
    \begin{subfigure}[h!]{0.48\textwidth}
        \centering
        \includegraphics[width=\linewidth]{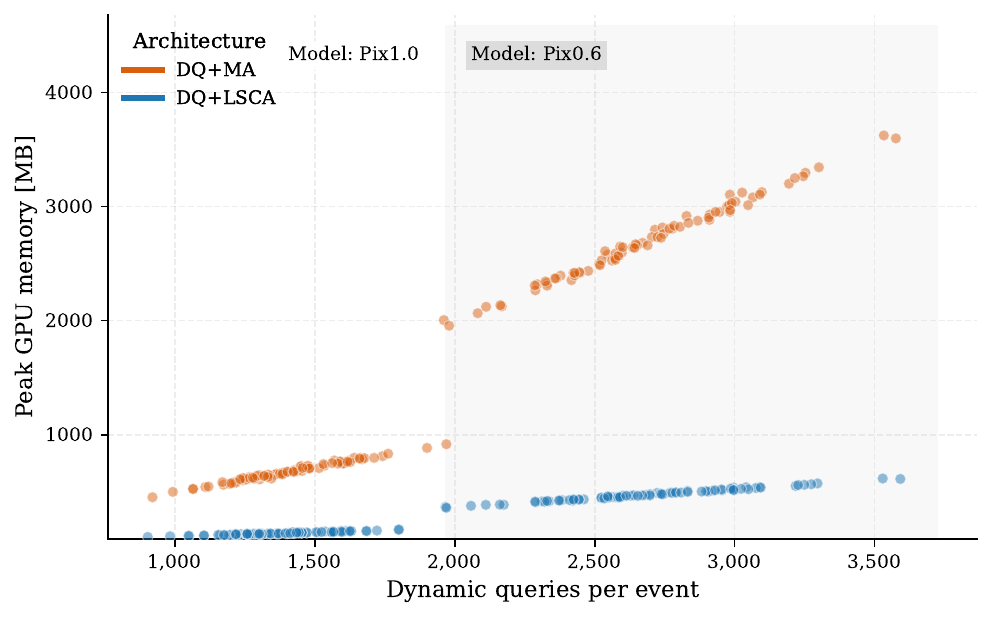}
        \caption{Peak inference GPU memory usage as a function of dynamic query count.}
        \label{fig:regime_memory_scaling_queries}
    \end{subfigure}
    \caption{
    Allocated inference GPU memory across pixel-only detector configurations of increasing complexity, shown as a function of retained hit multiplicity and dynamic query count.
    Colours distinguish the decoder variants, Masked Attention (DQ+MA) and Local Strided Cross-Attention (DQ+LSCA).
    Circular markers denote allocated memory measured for individual events, which reflects event-level scaling.
    The unshaded region indicates the Pix1.0 configuration regime while the shaded region indicates the Pix0.6 configuration.
    }
    \label{fig:regime_memory_scaling}
\end{figure*}

Figure~\ref{fig:regime_memory_scaling} shows the event-level scaling of peak allocated inference memory. Memory usage increases with event size for both attention formulations, but DQ+LSCA exhibits a weaker dependence and a lower absolute footprint than DQ+MA. As shown in Table~\ref{tab:systems_all}, the allocated-memory benefit of DQ+LSCA is largest during inference. Relative to DQ+MA, DQ+LSCA reduces peak allocated inference memory by approximately $80\%$ for both the Pix1.0 and Pix0.6 configurations. During training, the corresponding reductions are smaller but consistent, at approximately $20\%$ for the Pix1.0 and $10\%$ for the Pix0.6 configuration.

Together with the tracking results in Section~\ref{sec:model_a_tracking}, these measurements show that DQ+LSCA provides a favourable accuracy--efficiency trade-off for pixel-only reconstruction. Relative to masked attention, it retains high double-majority efficiency with some loss in hit-assignment fidelity, while substantially reducing inference latency and peak allocated inference memory. The corresponding reductions in peak allocated training memory are smaller, but remain consistent across both pixel-only regimes.

\subsection{Extension to Strip-Inclusive Tracking}
\label{sec:strip_results}

We finally evaluate the strip-inclusive configuration, PixStrip1.0, which extends the pixel-only setup by incorporating strip measurements. This is substantially more challenging than the pixel-only configurations, since the model must reconcile two detector subsystems with different spatial resolutions and geometries while predicting associations over longer per-track hit sequences. The extension is therefore not a pure scaling study in input size, but a test of whether the framework can support heterogeneous detector measurements.

Tracking performance for the strip-inclusive configuration is summarised in Table~\ref{tab:model_c_tracking}. The masked-attention decoder achieves high combined filter-plus-tracking double-majority efficiency with a fake rate of $0.3\%$ and a perfect-match efficiency of $93.5\%$. Replacing masked attention with DQ+LSCA reduces the double-majority efficiency by $1.8$ percentage points and increases the fake rate by $0.1$ percentage points. The degradation is larger for the stricter perfect-match criterion, which decreases by $6.3$ percentage points, indicating that exact hit assignment is more sensitive to the fixed local attention window in the heterogeneous pixel--strip setting.

\begin{table}[h]
\centering
\caption{
Tracking performance in the pixel--strip configuration, Model PixStrip1.0. Reported metrics are
the double-majority fake rate, the combined hit-filter plus tracking double-majority efficiency,
and the combined hit-filter plus tracking perfect-match efficiency.
}
\label{tab:model_c_tracking}
\begin{tabular}{lccc}
\toprule
Decoder & $f_{\mathrm{DM}}$ [\%] & $\epsilon^{\mathrm{filter+track}}_{\mathrm{DM}}$ [\%] & $\epsilon^{\mathrm{filter+track}}_{\mathrm{perfect}}$ [\%] \\
\midrule
DQ+MA   & 0.3 & 98.1 & 93.5 \\
DQ+LSCA & 0.4 & 96.3 & 87.2 \\
\bottomrule
\end{tabular}
\end{table}

As shown in Table~\ref{tab:systems_strip}, replacing masked attention with DQ+LSCA reduces the PixStrip1.0 configuration tracking-only inference latency from $101.9 \pm 30.2~\mathrm{ms}$ to $74.8 \pm 22.8~\mathrm{ms}$, corresponding to a reduction of approximately $27\%$. Peak allocated GPU memory is reduced by approximately $29\%$ during inference and by approximately $16\%$ during training. These allocated-memory reductions are smaller in relative terms than in the pixel-only occupancy-scaling study, consistent with the strip-inclusive architecture having a larger contribution from components not directly affected by the LSCA replacement, including the deeper encoder and detector-specific input and association heads. DQ+LSCA therefore reduces the decoder-side computational burden in the strip-inclusive regime while retaining high double-majority efficiency, although with a larger loss in exact hit-assignment fidelity.

\begin{table*}[h]
\centering
\caption{Computational performance for the strip-inclusive Model PixStrip1.0 configuration. 
`Trk Time' denotes tracking-only inference latency, while `HF + Trk Time' denotes end-to-end inference latency including the upstream hit filter (HF) and tracking model.
`Train Mem.' and `Infer. Mem.' denote peak allocated GPU memory during training and inference, respectively. Timing values are reported as the mean per event, with quoted spreads corresponding to the RMS over events. All memory values are GPU allocated-memory measurements.
}
\label{tab:systems_strip}
\begin{tabular}{lcccc}
\toprule
Decoder
& Trk Time [ms]
& HF + Trk Time [ms]
& Train Mem. [GB]
& Infer. Mem. [GB] \\
\midrule
DQ+MA   & $101.9 \pm 30.2$ & $150.2 \pm 32.2$ & 9.5 & 0.7 \\
DQ+LSCA & $74.8 \pm 22.8$ & $123.1 \pm 25.3$ & 8.0 & 0.5 \\
\bottomrule
\end{tabular}
\end{table*}

This study demonstrates that the dynamic-query and LSCA framework can be extended beyond pixel-only inputs to heterogeneous detector measurements. The results for the PixStrip1.0 configuration should be interpreted as a proof of principle for the tractability of query-based transformer tracking with strip measurements, rather than as a fully optimised strip-inclusive reconstruction benchmark. As described in Section~\ref{sec:model_variants}, the present architecture uses pixel hits to initialise and refine trajectory queries, while strip measurements are associated through detector-specific output heads. Alternative mechanisms for integrating pixel and strip information, such as unified association heads, separate encoder streams, or sequential pixel-to-strip extension, remain important directions for future study. Such architectural refinements will be important for maintaining reconstruction performance when extending the strip-inclusive configuration to lower transverse-momentum thresholds.

%% file: sections/7.conclusion.tex
\section{Conclusion}
\label{sec:conclusions}

We have presented a geometry-aware dynamic-query transformer architecture for trajectory reconstruction in high-multiplicity scientific sensor data. The model addresses two limitations of query-based transformer decoders in this regime. The first is the misalignment between fixed, input-independent queries and variable track multiplicities. The second is the cost of decoder cross-attention, which scales with both the number of trajectory queries and the number of detector hits. Static learned queries are replaced with event-dependent decoder inputs initialised from representative encoder-level hit embeddings, aligning the candidate set with physical trajectory hypotheses. Local Strided Cross-Attention (LSCA) then exploits the induced geometric ordering by replacing learned mask-gated cross-attention with a geometry-defined banded support, restricting attention to physically plausible query--hit interactions and exposing structured block sparsity for efficient sparse-attention execution.

The dynamic-query architecture improves reconstruction quality relative to the previous fixed-query baseline while enabling a substantially lighter decoder. In the Pix1.0 configuration, dynamic query initialisation, together with the associated architectural modifications, increases combined filter-plus-tracking double-majority efficiency by $4.0$ percentage points, reduces the fake rate by almost $60\%$, and increases perfect-match hit-assignment efficiency by $5.4$ percentage points. At the same time, the reduced decoder cost lowers tracking-only latency by approximately $40\%$ relative to the fixed-query baseline. These improvements extend the practical operating regime of the model. The dynamic-query architecture scales to $\pT \geq 0.6~\mathrm{GeV}$ over the studied $|\eta| \leq 4$ pixel acceptance, and is also extended to the strip-inclusive PixStrip1.0 configuration with heterogeneous detector inputs.

LSCA provides a complementary operating point for regimes in which inference latency and memory footprint are prioritised. By replacing learned mask-gated query--hit attention with local diagonal support implemented through block-sparse masks, LSCA substantially reduces decoder-side resource usage at the cost of some loss in hit-assignment fidelity. In the high-occupancy Pix0.6 configuration, DQ+LSCA reduces tracking-only latency by more than $50\%$ relative to DQ+MA and reduces peak allocated inference memory by up to a factor of $6$ in the pixel-only configurations. It also largely removes the observed dependence of decoder latency on retained hit multiplicity and dynamic-query count over the studied event range, indicating that the incremental cross-attention cost becomes subdominant to other timing overheads. In the strip-inclusive configuration, DQ+LSCA reduces tracking-only latency by approximately $27\%$, peak allocated inference memory by approximately $29\%$, and peak allocated training memory by approximately $16\%$ relative to DQ+MA, while retaining high combined double-majority efficiency.

The resulting reduction in model footprint improves the practical deployment characteristics of transformer-based tracking and makes it feasible to extend the reconstruction task to higher-occupancy regimes and heterogeneous detector inputs without prohibitive memory or runtime costs. Even for the largest configurations studied here, peak allocated inference memory remains below $4~\mathrm{GB}$, leaving substantial headroom on modern accelerator hardware for batched execution to improve effective throughput and reduce the amortised per-event latency. Future work should explore batched execution, alternative query-initialisation strategies, improved strip-hit representations, and integration with downstream track-fitting pipelines.

Taken together, these results show that the scalability limitations of query-based transformers for sparse reconstruction can be mitigated by using detector geometry to structure both decoder initialisation and cross-attention. More broadly, object queries need not be fixed learned parameters. Prior knowledge of the sensor layout can instead guide the selection of encoded input measurements that initialise an event-specific query set, allowing decoder capacity to adapt to the observed event while anchoring each query to a candidate physical object. The ordering induced by this construction can then be made computationally useful by translating physically plausible query--hit locality into a sparse attention pattern compatible with efficient block-sparse execution. Although demonstrated for charged-particle tracking, this approach applies more broadly to sparse reconstruction problems in which measurements admit a physically meaningful ordering coordinate and representative object seeds.